\documentclass[onecolumn,10pt]{IEEEtran}
\usepackage{cite}
\usepackage{amsmath,amssymb,amsfonts}
\usepackage{graphicx}
\usepackage{subcaption}
\usepackage{cleveref}
\usepackage{color, xcolor}
\usepackage{multirow}
\usepackage{pstricks}

\usepackage{amsthm}
\usepackage{booktabs}
\usepackage{mathrsfs} 
\usepackage{multirow}
\usepackage{algorithm}
\usepackage{algpseudocode}
\usepackage{caption}
\newtheorem{theorem}{Theorem}
\newtheorem{lemma}{Lemma}
\newtheorem{definition}{Definition}

\newtheorem{example}{Example}
\newtheorem{remark}{Remark}
\newtheorem{construction}{Construction}

\newcommand{\tabcaption}{\def\@captype{table}\caption}
\newcommand{\tabincell}[2]{\begin{tabular}{@{}#1@{}}#2\end{tabular}}

\begin{document}
	\title{A New Construction Structure on Multi-access Coded Caching with Linear Subpacketization: Cyclic Multi-Access Non-Half-Sum Disjoint Packing}
	
	\author{
		\IEEEauthorblockN{%
			Mengyuan
			Li\IEEEauthorrefmark{1},
			Minquan Cheng\IEEEauthorrefmark{1},
			Kai Wan \IEEEauthorrefmark{2},	
			Giuseppe Caire \IEEEauthorrefmark{3}
		}
		
		\IEEEauthorblockA{\IEEEauthorrefmark{1} Guangxi Normal University, 541004 Guilin, China, 
			mengyuanlli@163.com, chengqinshi@hotmail.com}
		\IEEEauthorblockA{\IEEEauthorrefmark{2} Huazhong University of Science and Technology, 430074  Wuhan, China, kai\_wan@hust.edu.cn}%
		\IEEEauthorblockA{\IEEEauthorrefmark{3}Technische Universit\"{a}t Berlin, 10587 Berlin, Germany,  caire@tu-berlin.de }
	}
	\maketitle
	\begin{abstract}
		We consider the $(K,L,M,N)$ multi-access coded caching system introduced by Hachem et al., which consists of a central server with $N$ files, $K$ cache nodes and $K$ cacheless users, where each user can access $L$ cache nodes in a cyclic wrap-around fashion. At present, several existing schemes achieve competitive transmission load, but their subpacketization levels grow exponentially with the number of users. In contrast, schemes with linear or polynomial subpacketization always incur higher transmission loads. We aim to design a multi-access coded caching scheme with linear subpacketization $F$ while maintaining low transmission load. Recently, Cheng et al. proposed a construction framework for coded caching schemes with linear subpacketization (i.e., $F=K$) called non-half-sum disjoint packing (NHSDP). Inspired by this structure, we introduce a novel combinatorial structure named cyclic multi-access non-half-sum disjoint packing (CMA-NHSDP) by extending NHSDP to MACC system. By constructing CMA-NHSDP, we obtain a new class of multi-access coded caching schemes. Theoretical and numerical analyses show that our scheme achieves lower transmission loads than some existing schemes with linear subpacketization. Moreover, the proposed schemes achieves lower transmission load compared to existing schemes with exponential subpacketization in some case.
	\end{abstract}

	\section{Introduction}
	
	With the popularity of Internet-connected devices, wireless networks have faced increasing traffic pressure during peak service hours. Furthermore, the highly time-varying nature of network traffic often leads to congestion during peak hours and underutilized resources during off-peak hours. Caching is an effective technique to solve this problem, which pre-stores popular contents at network cache nodes during off-peak periods and exploits local cache resources to obtain ``local caching gain'', thereby reducing the overall network traffic\cite{BBD}. Maddah-Ali and Niesen (MN) \cite{MN} proposed the first coded caching scheme, which achieves a ``global caching gain" by designing each broadcast message to simultaneously send multiple users.
	
	
	In the $(K,M,N)$ centralized caching system introduced by Maddah-Ali and Niesen, a central server storing $N$ files of equal size is connected to $K$ users, each equipped with a local cache of size $M$ files. The server communicates with the users over a  error-free shared-link. A $(K,M,N)$ coded caching scheme consists of two phases: the placement phase and the delivery phase. In the placement phase, each file is divided into $F$ equal-length data packets, some of which are stored in each user's cache without knowledge of future demand. The quantity $F$ is referred to as the subpacketization. In the delivery phase, each user randomly requests a file from the server. Based on these requests and the cached content of each user, the server broadcasts coded messages via an error-free shared link to ensure that each user can recover the requested file. The transmission load $R$ is defined as the worst-case normalized transmission load across all possible user demands, and the design objective of the scheme is to minimize $R$.

	The MN scheme employs an uncoded placement strategy\footnote{Original files are stored in the user cache directly; otherwise, it would be a coded placement strategy.} combined with a coded delivery strategy. Under the constraint of uncoded placement, it has been shown that the MN scheme achieves the optimal transmission load when $K \leq N$ \cite{WTP2016,WTP2020}. For the case $K > N$, Yu et al. \cite{YMA2018} further optimized the scheme by eliminating redundant transmissions in the MN scheme, generating an optimal uncoded placement scheme in that regime. Therefore, the MN scheme has been extensively adapted to various network topologies, including but not limited to cache-aided relay networks~\cite{Karim2020Cache}, heterogeneous networks~\cite{Li2019Coded}, multi-access networks with cyclic wrap-around topology~\cite{HKD}.

	However, subpacketization of the MN scheme increases exponentially with the number of users $K$, which severely limits its practicality. Therefore, subsequent research has focused on reducing subpacketizations. The grouping method \cite{SJTLD,CJWY} is highly effective for reducing subpacketization, but this reduction often comes at the cost of a significantly increased transmission load. To further propose a coded caching scheme with low subpacketizations, the study in \cite{YCTC} introduced a combinatorial structure known as the Placement Delivery Array (PDA), which characterizes coded caching schemes employing uncoded placement and one-shot delivery. Since then, numerous construction schemes based on PDA  with reduced subpacketization have been proposed, including \cite{SZG,YTCC,CJWY,CJYT,ASK,CJTY,CWZW,WCWC2022ISIT}. However, a PDA can only describe the placement and delivery phases in shared-link model, without providing direct guidance for constructing coded caching schemes. Recently, \cite{CWWC} proposed a novel combinatorial structure called Non-Half-Sum Disjoint Packing (NHSDP), which unifies placement and delivery strategies into a single combinatorial condition. By NHSDPs, a class of coded caching schemes with linear subpacketization can be obtained, which achieves significantly lower, or only slightly higher transmission loads compared with some existing schemes under certain system parameters.
	
	\subsection{Multi-access Coded Caching}
	
	Caching at wireless edge nodes plays a crucial role in modern wireless systems, simultaneously boosting spatial and spectral efficiency and reducing communication costs~\cite{LCYM,SGDM}. Edge caches support the storage of diverse Internet-based content, including web resources, video streams, and software updates. In contrast to end-user-caches, which are constrained by the storage of individual devices and designed to serve only a single user, edge caches provide substantially larger storage capacities and can simultaneously support multiple users. These caches are deployed at local helper nodes, mobile edge computing (MEC) servers, and hotspots. The distinction is crucial, as content libraries in practical systems often far exceed the storage capabilities of mobile devices. 
	This paper studies an ideal coded edge caching framework with a linear multi-access topology, called as the multi-access caching model (illustrated in Fig.~\ref{multiaccess-system}), which was originally introduced in \cite{HKD}. Unlike conventional shared-link caching systems, the multi-access model consists of $K$ cache nodes, each of which can store $M$ files and is connected to $K$ cache-less users. Each user can access $L$ neighboring cache-nodes in a cyclic wrap-around manner while simultaneously receiving broadcast transmissions from the central server. According to the assumption in \cite{HKD} that accessing cache nodes will not generate load cost, this analysis only considers the broadcast load of the server. The rationality of this assumption lies in that the access of cache-nodes can be offloaded to an independent network infrastructure.
	
	
	Under the constraint of uncoded cache placement, the authors of \cite{HKD} introduced a multi-access coded caching scheme. They proved that the gap between its achievable transmission load and the information-theoretic lower bound is no more than $cL$, where $c$ is a constant.  Reference \cite{RK} proposed a new scheme using index coding, which achieves a lower transmission load than \cite{HKD} when $K < \frac{KLM}N + L$. Furthermore, for the case where $L \geq K/2$ and $N \geq K$, the authors of \cite{RK}  derived a converse bound under uncoded placement. They proved that the scheme has order optimality within a factor of 2. A transformation method was proposed in \cite{CWLZC} to adapt the MN scheme to the multi-access scenario. This adaptation achieves a lower transmission load than \cite{HKD} when $L \nmid K$. The work in \cite{RK2021} decomposed the multi-access caching problem into structured index coding problem. Based on solutions to the structured index coding problem, a new scheme has been proposed with a transmission load that does not exceed the transmission loads of the schemes in \cite{RK} and \cite{CWLZC}. It is worth noting that subpacketization of the above scheme increases exponentially with $K$.\cite{WCWL} proposed a consecutive cyclic placement strategy. By constructing a class of PDAs under this strategy, a novel multi-access coded caching scheme was introduced. This scheme has a subpacketization increasing linearly with the number of users and attains a minimal transmission load of $\frac{K-KML/N}{2 \left\lfloor \frac{K}{K - KML/N + 1} \right\rfloor + 1}$. 
	All the aforementioned schemes are designed for general system parameters. A summary is provided in Table \ref{tab-knowschemes}. For specific parameter configurations, several schemes with reduced subpacketization have been introduced in \cite{SPE,SR_arX2021,SR,MR}. These are also included in Table \ref{tab-knowschemes}.
	
	\begin{table}[htbp!]
		\centering
		\renewcommand{\arraystretch}{2.5}
		\setlength\tabcolsep{0.8pt} 
		\caption{Known multi-access coded caching schemes with uncoded placement.Where $K$ is the number of users, the number of cache nodes is equal to $K$, $L$ is the access degree and the cache-node memory ratio is $\frac{t}{K}$. If the range of $t$ is not specified in the Limitation section, then $t\in \{0,\ldots,\lfloor\frac{K}{L}\rfloor\}$.} \label{tab-knowschemes}
		\begin{tabular}{|c|c|c|c|}
			\hline
			Schemes&Subpacketization &Rate&Limitation  \\
			\hline
			\multirow{2}{*}{HKD scheme \cite{HKD}} & $L{K/L\choose t}$ & $\frac{K-t L}{1+t}$ & $L|K$ \\ \cline{2-2} \cline{3-3} \cline{4-4} & ${K\choose t}$& $\frac{K-t}{1+t}$ & $L\nmid K$ \\
			\hline
			
			\tabincell{c}{RK1 scheme  \cite{RK}} & ${K-t(L-1)-1 \choose t-1}\frac{K}{t}$ & $\frac{(K-t L)^{2}}{K}$ & \\ \hline

			\tabincell{c}{CW scheme  \cite{CWLZC}} & ${K-t(L-1) \choose t}K$ & $\frac{K-t L}{1+t}$ & \\ \hline
			
			\tabincell{c}{RK2 scheme  \cite{RK2021}} & ${K-t(L-1)-1 \choose t-1}\frac{K}{t}$& $ \frac{\sum_{b \in B} \min \{2(K - t L) + t -1- \hat{b} , K\}}{|\hat{S}|(t + 1)}$& \\ \hline
			
			\multirow{3}{*}{WCWL scheme \cite{WCWL}} 
			& $K$ & $ \frac{(K-tL)(K-tL+1)}{2K} $ & $(K-tL+1)|K\ \text{or}\ K-tL=1$\\ \cline{2-4}
			& $\left(2\lfloor\frac{K}{K-tL+1}\rfloor+1\right)K$ & $\frac{K-tL}{2\lfloor\frac{K}{K-tL+1}\rfloor+1}$ & $\langle K\rangle_{K-tL+1}=K-tL$ and $K-tL>1$\\ \cline{2-4}
			& $2\lfloor\frac{K}{K-tL+1}\rfloor K$ & $\frac{K-tL}{2\lfloor\frac{K}{K-tL+1}\rfloor}$ & $\mbox{otherwise}$\\ \cline{2-4}
			\hline 
			
			\tabincell{c}{SPE scheme \cite{SPE}} & $\frac{K(K-2L+2)}{4}$ & \tabincell{c}{$\geq \frac{K-2L}{4}$} &$4|K(K-2L+2)$, $t=2$\\
			\hline
			
			\multirow{3}{*}{SR1 scheme  \cite{SR_arX2021}} 
			&  
			& $ \frac{1}{K} $ & $ K - tL = 1; \text{$t$, $K$ are coprime} $\\ \cline{3-4} &\tabincell{c}{$\leq K^2$} &$2\sum_{r=\frac{K-tL}{2}+1}^{K-tL}\frac{1}{1+\lceil\frac{tL}{r}\rceil} $ & $ K - tL \text{ is even}; \text{$t$, $K$ are coprime} $\\ \cline{3-4}  & &$\frac{1}{\left(\lceil\frac{2tL}{K-tL+1}\rceil+1\right)} + \sum_{r=\frac{K-tL+3}{2}}^{K-tL}\frac{2}{1+\lceil\frac{tL}{r}\rceil} $ & $ K - tL > 1 \text{ is odd}; \text{$t$, $K$ are coprime} $\\ \cline{3-4}
			\hline
			
			\tabincell{c}{SR2 scheme  \cite{SR}} & $K$ & $\frac{(K-t L)(K-t L+t)}{2K}$&$t|K$, $(K-t L+t)|K$\\
			\hline
			MR scheme  \cite{MR} & $K$ & $\geq\left\lceil \frac{K(K-L)}{2+2\lfloor \frac{L}{K-L+1} \rfloor} \right\rceil \frac{1}{K}$& $t=1$ \\ \hline

			
		\end{tabular}
	\end{table}

	Furthermore, the work in \cite{NR2021} introduces a multi-access coded caching scheme that incorporates secure delivery mechanisms. When the additional memory allocated for security purposes is excluded, this scheme achieves the same transmission load as the RK1 scheme presented in \cite{RK}, and the required subpacketization level does not exceed $K$. In \cite{NR}, a converse bound based on the cut-set argument was established, and a new scheme of coded placement was proposed for the memory regime satisfying $M < (N - K + L)/K$. This scheme is proven to be optimal under the condition $N \leq K$. Several studies \cite{KMR, MR2021, DR} have investigated multi-access coded caching systems with a slightly modified topology based on cross-resolvable designs. The authors of \cite{OG, ZWCC} explored a two-dimensional multi-access scenario, where the mobile users move on a two-dimensional grid and can access the nearest cache-nodes inside the grid. The work in \cite{CWEC} presents a novel scheme for a general MACC setting, unifying existing MACC schemes with the customized solution for specific encoding cache scenarios. This scheme enhances user distribution density while maintaining good load performance.
	
	As observed in Table~\ref{tab-knowschemes}, while the schemes in \cite{HKD, RK, CWLZC, RK2021} achieve competitive transmission loads, their subpacketization grows exponentially with the number of users. In contrast, \cite{WCWL, SR, MR} and other schemes maintain linear or polynomial subpacketization, but at the cost of limited coded caching gains and higher transmission loads. This trade-off reveals a key limitation in multi-access coded caching. Although the PDA serves as a combinatorial structure that characterizes placement and delivery strategies, its definition alone does not directly yield constructions that simultaneously achieve low subpacketization and significant coded caching gain. Therefore, constructing a general framework for PDA that guarantee linear subpacketization while maintaining near-optimal transmission performance is an important and meaningful research direction.

	\subsection{Research motivation and contribution}
	In this paper, we focus on designing multi-access coded caching schemes with linear subpacketization, aiming to minimum the transmission load under uncoded placement and one-shot delivery strategies. Through the analysis of Table~\ref{tab-knowschemes}, we can find that existing multi-access coded caching schemes with linear or polynomial subpacketization still face challenges in balancing subpacketization and rate. Motivated by the structure of non-half-sum disjoint packing (NHSDP) in shared-link caching systems, we extend this interesting structure to the multi-access network and propose a novel structure called cyclic multi-access non-half-sum disjoint packing (CMA-NHSDP). This structure offers a unified combinatorial framework for designing multi-access coded caching schemes with linear subpacketization. More specifically, the main contributions are summarized as follows:
	\begin{itemize}
		\item We introduce a new class of PDAs, called $L$-continuous $(K,F,Z,S)$ PDAs, which is used 
		for characterizing a $F$-division $(K,L,M,N)$ MACC scheme with memory ratio $\frac{M}{N}=\frac{Z}{LF}$, subpacketization $F$, and transmission load $R=\frac{S}{F}$.
		\item  We propose a novel combinational structure termed CMA-NHSDP. It is worth noting that the NHSDP proposed in \cite{CWWC} for shared-link networks cannot be directly applied to MACC systems, as each user in the shared-link system is equipped with independent caches. The structure CMA-NHSDP is a generalized framework for extending NHSDP to MACC systems.
		Given an $(L, v, g, b)$ CMA-NHSDP for any odd positive integers $v$, any positive integers $L$, $g$ and $b$, we can obtain an $L$-continuous $(K=v,F=v,Z=v-bg,S=bv)$ PDA. When $K$ is even, we can add one virtual user to construct a scheme based on the $(L, K+1, g, b)$ CMA-NHSDP.
		\item  We propose a CMA-NHSDP construction framework by introducing an offset function into the NHSDP construction. This offset function is designed to ensure that the differences between sorted integers within our proposed structure are $kL+1(k\in \mathbb{N})$, while maximizing the number of integers, i.e., minimizing the memory ratio. Based on the CMA-NHSDPs, we can obtain a $(K=v,L,M,N)$ multi-access coded caching scheme with the memory ratio $M/N=\frac{v-2^n\prod_{i=1}^{n}m_i}{vL}$, coded caching gain $g=2^n$, and transmission load $R=\prod_{i=1}^{n}m_i$ for any odd integer  $v\geq \prod_{i=1}^{n}(1+2m_i)+2\sum_{i=1}^{n}\left(g(i)\prod_{j=i+1}^{n}(1+2m_j)\right)$  where $n$ and $m_1$, $m_2$, $\ldots$, $m_n$ are any positive integers. In particular, when $\frac{m_1}{L}=m_2=\cdots=m_n=\left\lfloor \frac{(\frac{v}{L})^{1/n}-1}{2} \right\rfloor$, we obtain a $(K=v,L,M,N)$ multi-access coded caching scheme with the memory ratio $\frac{M}{N}=\frac{1}{L}-\frac{\left\lfloor(\frac{v}{L})^{\frac{1}{n}}-1 \right\rfloor^{n}}{v}$, subpacketization $F=K$, coded caching gain $g=2^n$, and  transmission load $R={\left\lfloor \frac{v^{1/n}-1}{2} \right\rfloor}^nL$.
		
		\item  Theoretical and numerical comparisons show that our scheme achieves lower transmission load compared to \cite{SPE,SR,SR_arX2021,MR} with linear or polynomial subpacketization. Furthermore, it achieves significantly reduced subpacketization and slightly improved transmission load than schemes in \cite{HKD,RK,RK2021,CWEC} with exponential subpacketization. 
	\end{itemize}
	
	\subsection{Organization and notations}
	The remainder of this paper is organized as follows. Section~\ref{sec-prelimin} introduces the multi-access coded caching system model, reviews the concept of placement delivery arrays (PDA) and non-half-sum disjoint packing (NHSDP). In Section~\ref{sec-CMA-NHSDP}, we introduce the cyclic multi-access non-half-sum disjoint packing (CMA-NHSDP) and show that it can be used to generate an $L$-continuous PDA. Section~\ref{sec-Construct-CMA-NHSDP} introduce a construction framework of CMA-NHSDPs. The performance analysis of the proposed scheme is provided in Section~\ref{sec-perf-ana}. Finally, Section~\ref{sec-conclu} concludes this paper.

	Notations: In this paper, we will use the following notations. Let bold capital letter, bold lowercase letter, and curlicue letter denote array, vector, and set respectively; let $|A|$ denote the cardinality of the set $A$; define $[a] = \{1, 2, \ldots, a\}$ and $[a : b]$ is the set $\{a, a+1, \ldots, b-1, b\}$; $\lfloor a \rfloor$ denotes the largest integer not greater than $a$. $a|b$ denotes $a$ divides $b$; $a\nmid b$ denotes $a$ does not divide $b$. We define that $\langle K \rangle_t = K \bmod t$; $\mathbb{Z}_v$ is the ring of integer residues modulo $v$; $\mathcal{B} \setminus \mathcal{A}$ denotes the set consisting of elements that belong to set $\mathcal{B}$ but not to set $\mathcal{A}$; For any array $\mathbf{P}$, $\mathbf{P}(i,j)$ denotes the entry in the $i^{th}$ row and $j^{th}$ column.

	\section{PRELIMINARIES}\label{sec-prelimin}
	In this section, we will review a multi-access coded caching system, placement delivery array, and non-half-sum disjoint packing.
	\subsection{Multi-access coded caching system}\label{system model}
\begin{figure}[ht]
		\centering
		\includegraphics[width=5in]{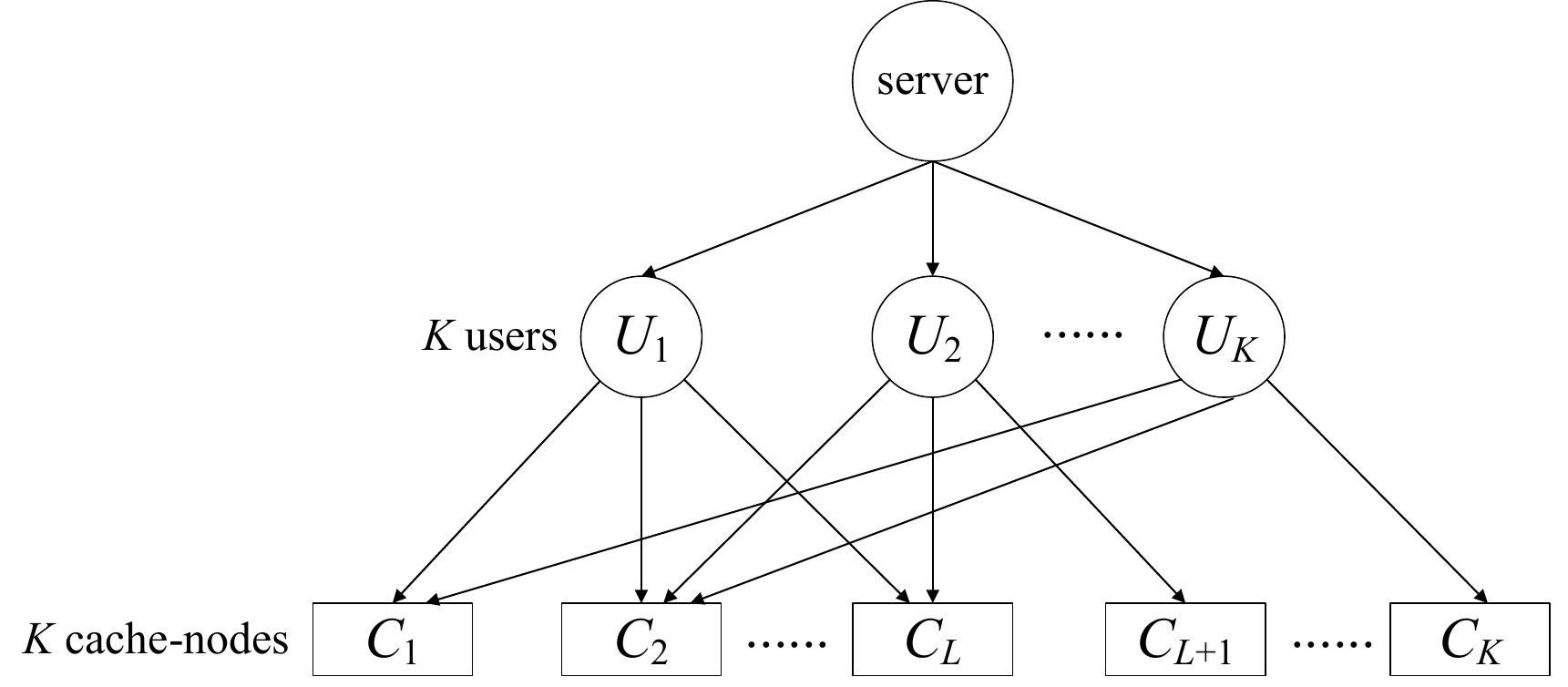}
		\vskip 0.2cm
		\caption{Multi-access coded caching system.}\label{multiaccess-system}
	\end{figure}

	A $(K,L,M,N)$ multi-access coded caching (MACC) system \cite{HKD} (Fig.~\ref{multiaccess-system}) comprises a central server with a library $\mathcal{W} = \{\mathbf{w}_n \mid n\in [N]\}$ of $N$ unit-sized files, connected to $K$ cache-less users (denoted by $U_1, U_2, \ldots, U_k$) via an error-free shared link. The system also includes $K$ cache-nodes (denoted by $C_1, C_2, \ldots, C_k$), each with a cache of size $M$ files where $0 \leq M \leq N$. Each user can access multiple groups of $L$ consecutive cache-nodes in a cyclic wrap-around manner.
	
	A $F$-division $(K,L,M,N)$ multi-access coded caching scheme contains two independent phases:
	
	$\bullet$ {\bf Placement phase:} The server divides each file $\mathbf{w}_n$ into $F$ equal-sized packets, denoted by $\mathbf{w}_n=\{\mathbf{w}_{n,j}\ | j\in[F]\}$ where $n\in[N]$, and places some packets into each cache-node. The packets stored in cache-node $C_k$ are denoted by $\mathcal{Z}_{C_k}$. Each user can access packets from its connected cache-nodes, and the total size of these different contents is defined as the \emph{local caching gain}, which ideally should be as large as possible. The placement phase is done without prior knowledge of user demands.
	
	$\bullet$ {\bf Delivery phase:} Each user randomly requests a file $\mathbf{w}_{d_k}$, where $d_k \in [N]$, forming the request vector $\mathbf{d} = (d_1, \dots, d_K)$. Based on ${\bf d}$ and the contents stored in the cache-nodes, the server broadcasts at most $S_{\mathbf{d}}$ coded packets, each of which is linearly combined by the required packets to the users, such that each user can decode its requested file. 
	
	The {\em transmission load} (or rate) is defined as the worst case transmission amount $R=\max_{{\bf d} \in [1:N]^K} S_{\mathbf{d}}/F$. Our objective is to design a scheme such that the transmission load is as small as possible given the parameters $K$, $L$, $M$ and $N$.

	
	The authors in \cite{CWLZC} introduced the following triple of arrays—node-placement, user-retrieve, and user-delivery to characterize the placement strategy for cache nodes, the retrieve content statuses of users, and the delivery strategy of server, respectively. 
	\begin{definition}[Node-placement, user-retrieve, and user-delivery arrays, \cite{CWLZC}]\rm
		\label{defn:three arrays}
		\
		\begin{itemize}
			\item An $F \times K$ node-placement array $\mathbf{C}$ consists of stars and nulls, where $F$ and $K$ represent the subpacketization and the number of cache-nodes,  respectively. The entry  located at the position $(j,k)$ in  $\mathbf{C}$ is star if and only if the $k^{\text{th}}$ cache-node   caches  the $j^{\text{th}}$ packet of each $\mathbf{w}_n$ where $n\in [N]$. 
			\item An $F \times K$ user-retrieve array $\mathbf{U}$ consists of stars and nulls, where $F$ and $K$ represent the subpacketization and the number of users, respectively. The entry at the position $(j,k)$ in  $\mathbf{U}$
			is star if and only if the $k^{\text{th}}$ user can retrieve   the $j^{\text{th}}$ packet  of each $\mathbf{w}_n$ where $n\in [N]$.
			\item An $F \times K$ user-delivery array $\mathbf{Q}$ consists of $\{*\}\cup[S]$, where $F$, $K$ and the stars in $\mathbf{Q}$ have the same meaning as $F$, $K$ of $\mathbf{U}$ and the stars in $\mathbf{U}$, respectively. Each integer represents a multicast message, and $S$ represents the total number of multicast messages transmitted in the delivery phase.
			\hfill $\square$
		\end{itemize}
	\end{definition}
	By Definition~\ref{defn:three arrays}, we can obtain a MACC scheme by constructing $\mathbf{C}$, $\mathbf{U}$, and $\mathbf{Q}$. In particular, when the access degree $L=1$, the node-placement array is equivalent to the user-retrieve array and has the same star positions of the user-delivery array. This implies that each cache node can be regarded as user's individual steerage device. In this case, a MACC scheme is exactly the scheme for the dedicated shared link network scenario. In fact, the authors in \cite{YCTC} proposed the following interesting combinatorial structure to characterize the scheme under uncoded placement and one-shot delivery strategy.  
	\subsection{Placement delivery array}
	
	\begin{definition}[PDA\cite{YCTC}]\rm
		\label{def-PDA}
		For positive integers $K, F, Z$ and $S$, an $F \times K$ array $\mathbf{P} = (p_{i,j})$, $i \in [0, F)$, $j \in [0, K)$ with alphabet set $\{*\}\cup [S]$ is called a $(K, F, Z, S)$ placement delivery array (PDA) if it satisfies the following three conditions:
		\begin{itemize}
			\item[C$1$.] the symbol $*$ occurs exactly $Z$ times in each column;
			\item[C$2$.] each integer occurs at least once in the array;
			\item[C$3$.] for any two distinct entries $p_{i_1,j_1}$ and $p_{i_2,j_2}$, such that $p_{i_1,j_1} = p_{i_2,j_2} = s$ is an integer only if
			\begin{itemize}
				\item[a)] $i_1 \neq i_2$, $j_1 \neq j_2$, i.e., they lie in distinct rows and distinct columns;
				\item[b)] $p_{i_1,j_2} = p_{i_2,j_1} = *$, i.e., the corresponding $2 \times 2$ sub - array formed by the rows $i_1$, $i_2$ and the columns $j_1$, $j_2$ must be of the following form
				$$
				\begin{pmatrix}
					s & * \\
					* & s
				\end{pmatrix}
				\quad \text{or} \quad
				\begin{pmatrix}
					* & s \\
					s & *
				\end{pmatrix}.
				$$
			\end{itemize}
		\end{itemize}
	\end{definition}

	The authors in \cite{CWLZC} also pointed out that if the user-delivery array $\mathbf{Q}$ satisfies the conditions C$2$ and C$3$, each user can decode its required file. That is the following statement. Specifically, if $\mathbf{Q}(j,k)=s$ is an integer, it implies that the $j^{\text{th}}$ packet of all files is not accessible to user $U_k$, and the server broadcasts a multicast message (i.e., the XOR of all the requested packets indicated by $s$) to all the users at time slot $s$. Condition C$3$ of Definition \ref{def-PDA} ensures that each user can get its desired packet, since all the other packets in the multicast message are accessible to it. The occurrence number of integer $s$ in $\mathbf{Q}$, denoted by $g_s$, is called the coded caching gain of the PDA in time slot $s$, since the message broadcast at time slot $s$ can serve $g_s$ users simultaneously. Condition C2 of Definition \ref{def-PDA} implies that the number of messages broadcasted by the server is exactly $S$, so the transmission load is $R=S/F$.

	\begin{example}\rm
		\label{exam-three-array}
		We take the following $4 \times 4$ array as the node-placement array $\mathbf{C}$ to generate a $F$-division $(K=4,L=2,M,N)$ MACC scheme. 
		$$\mathbf{C}=\begin{pmatrix}
			&*& & \\
			& &*& \\
			& & &*\\
			*& & & \\
		\end{pmatrix}$$
		\begin{itemize}
			\item \textbf{Placement Phase:} By Definition \ref{defn:three arrays}, the server divides each file $\mathbf{w}_n$ where $n\in[4]$ into $F=4$ packets, i.e., $\mathbf{w}_n=\{\mathbf{w}_{n,j} | j\in[4]\}$. Recall that each entry $\mathbf{C}(f,k) = *$ where $f\in[F]$ and $k\in[K]$ indicates that cache node $C_k$ stores the $f^{th}$ packet of   file. Then the cache nodes cache the following packets respectively
			\begin{align}
				\mathcal{Z}_{C_1}&=\{\mathbf{w}_{n,4} |\ n\in [4]\},\ \
				\mathcal{Z}_{C_2} =\{\mathbf{w}_{n,1} |\ n\in [4]\}, \nonumber\\
				\mathcal{Z}_{C_3}&=\{\mathbf{w}_{n,2} |\ n\in [4]\},\ \  \mathcal{Z}_{C_4} =\{\mathbf{w}_{n,3} |\ n\in [4]\}.\label{eq-exa-design-cache-node}
			\end{align}
			
			Since each user $U_k$ accesses $L=2$ consecutive cache nodes in a cyclic wrap-around manner, the packets accessed by users can be written as follows,
			\begin{align}
				\mathcal{Z}_{U_1}&=\{\mathbf{w}_{n,1}, \mathbf{w}_{n,4} |\ n\in [4]\},\ \
				\mathcal{Z}_{U_2} =\{\mathbf{w}_{n,1}, \mathbf{w}_{n,2} |\ n\in [4]\}, \nonumber\\
				\mathcal{Z}_{U_3}&=\{\mathbf{w}_{n,2}, \mathbf{w}_{n,3} |\ n\in [4]\},\ \  \mathcal{Z}_{U_4} =\{\mathbf{w}_{n,3}, \mathbf{w}_{n,4} |\ n\in [4]\}.\label{eq-exa-design-user-retrieve}
			\end{align} 
			So the user-retrieve array $\mathbf{U}$ is obtained as follows,
			$$\mathbf{U}=\begin{pmatrix}
				*&*& & \\
				&*&*& \\
				& &*&*\\
				*& & &*\\
			\end{pmatrix}.$$
			
			Next, we need to construct the user-delivery array $\mathbf{Q}$ for the delivery strategy. Recall that the stars in the user-retrieve array $\mathbf{U}$ and in the user-delivery array $\mathbf{Q}$ represent the same meaning. So we first set $\mathbf{Q} = \mathbf{U}$. By putting the integers in all the null entries of $\mathbf{Q}$, we can obtain the following array.
			$$\mathbf{Q}=\begin{pmatrix}
				*&*&1&4\\
				1&*&*&2\\
				3&2&*&*\\
				*&4&3&*\\
			\end{pmatrix}.$$ We can check that all the integers in $\mathbf{Q}$ satisfy Conditions C2 and C3 in Definition~\ref{def-PDA}. According to Definition \ref{defn:three arrays}, we can obtain the following delivery strategy by $\mathbf{Q}$. 
			
			\item \textbf{Delivery Phase:} Assume the user demand vector is $\mathbf{d} = (1,2,3,4)$. Recall that if the integer entry $\mathbf{Q}(f,k)=s$ where $f\in[F]$ and $k\in[K]$, the user $k$ can not retrieve the required packet $\mathbf{w}_{d_k,f}$. For instance, the entries $\mathbf{Q}(2,1)=\mathbf{Q}(1,3)=1$ imply that users $1$ and $3$ require the packets $\mathbf{w}_{1,2}$ and  $\mathbf{w}_{3,1}$, respectively. Therefore, the server constructs multicast message based on the demand vector and the user-retrieved packets as follows.
			
			\begin{align}
				X_1 &= \mathbf{w}_{3,1} \oplus \mathbf{w}_{1,2},\ \
				X_2 = \mathbf{w}_{4,2} \oplus \mathbf{w}_{2,3}, \nonumber\\
				X_3 &= \mathbf{w}_{1,3} \oplus \mathbf{w}_{3,4},\ \  X_4 = \mathbf{w}_{4,1} \oplus \mathbf{w}_{2,4}.\label{eq-exa-design-user-delivery}
			\end{align}
			According to multicast messages and user-retrieved packets, each user can decode the required content. For instance, at time slot $1$, $U_1$ recovers $\mathbf{w}_{1,2}$ from the received message $X_1$ because it can access to $\mathbf{w}_{3,1}$. Similarly, all users can decode their requested packets. The number of transmissions is $S=4$, leading to the load $R = S/F = 4/4 = 1$.\\
		\end{itemize}
	\end{example}
	
	By the aforementioned explanation of the relationship between user-delivery array $\mathbf{Q}$ and delivery strategy in Example \ref{exam-three-array}, we can obtain an MACC scheme by constructing a user-delivery array $\mathbf{Q}$ that satisfies conditions C$2$ and C$3$ in Definition \ref{def-PDA} and the following condition:
	\begin{itemize}
		\item	C$4$.     In each row, the symbol * appears in exactly $\frac{F-Z}{L}$ disjoint cyclic consecutive runs, each of length $L$.
		
		\hfill $\square$
	\end{itemize}	
	\begin{remark}\rm
		\label{con-c4}
		A $(K,F,Z,S)$ PDA \textbf{P} is called an $L$-continuous $(K,F,Z,S)$ PDA $\textbf{Q}$ if it satisfies C$4$. 
	\end{remark}
	\begin{lemma}\rm
		\label{th-Fundamental}
		An $L$-continuous $(K,F,Z,S)$ PDA $\textbf{Q}$ can generate a $F$-division $(K,L,M,N)$ MACC scheme with the cache-node memory size $M/N=Z/LF$, subpacketization $F$ and  transmission load $R=S/F$.
	\end{lemma}

	To construct an $L$-continuous $(K,F,Z,S)$ PDA $\textbf{Q}$ with linear subpacketization, we extend the non-half-sum disjoint packing (NHSDP) combinatorial structure in \cite{CWWC} to the MACC system. We first recall this structure in the following.
	\subsection{Non-half-sum disjoint packing}
	\begin{definition}[NHSDP \cite{CWWC}]\label{def-NHSDP}\rm
		For any positive odd integer $v$, a pair $(\mathbb{Z}_v,\mathfrak{D})$ where $\mathfrak{D}$ consists of $b$ $g$-subsets of $\mathbb{Z}_v$ is called $(v,g,b)$ non-half-sum disjoint packing if it satisfies the following conditions.
		\begin{itemize}
			\item The intersection of any two different blocks\footnote{In a NHSDP $(\mathbb{Z}_v, \mathfrak{D})$, each subset $\mathcal{D} \in \mathfrak{D}$ is called a block} in $\mathfrak{D}$ is empty;  
			\item For each $\mathcal{D}\in \mathfrak{D}$, the half-sum of any two different elements in $\mathcal{D}$ (i.e., the sum of the two elements divided by $2$\footnote{The operations are in $\mathbb{Z}_v$ and since $v$ is odd, then $2$ has an inverse, that is, $1/2$ in $\mathbb{Z}_v$ is also an element of $\mathbb{Z}_v$.}) does not appear in any block of $\mathfrak{D}$.
		\end{itemize} \hfill $\square$
	\end{definition}
	Let us take the following example with $v=5$ to further explain the concept of NHSDP. 
	\begin{example}\rm
		\label{exam-def-nh}
		When $v=5$ and $\mathfrak{D} = \{D_1 = \{1,4\},\; D_2 = \{2,3\}\}$, we have a pair $(\mathbb{Z}_5,\mathfrak{D})$. Since $D_1 \cap D_2 = \emptyset$, the first condition in Definition~\ref{def-NHSDP} holds.
		The half-sums of any two different elements in each block in $\mathfrak{D}$ are as follows, 
		\begin{equation}
			\begin{aligned}
				& {\rm in}\  \mathcal{D}_1:  \frac{1+4}{2}=0,\  
				&{\rm in}\  \mathcal{D}_2: \frac{2+3}{2}=0,\  
				\label{eq:ex D1 D2}
			\end{aligned}
		\end{equation}We have $\{0\}\cap\mathcal{D}_1=\{0\}\cap\mathcal{D}_2=\emptyset$, i.e., the second condition in Definition~\ref{def-NHSDP} holds. So $(\mathbb{Z}_5,\mathfrak{D})$ is a $(5,2,2)$ NHSDP.
	\end{example}
	Using an NHSDP, we can obtain a PDA with $F=K$ by the following construction. 
	\begin{construction}[\cite{CWWC}]\rm\label{cons-PDA-NHSDP}
		Given a $(v,g,b)$ NHSDP $(\mathbb{Z}_v,\mathfrak{D})$, then a $v\times v$ array $\mathbf{P}(f,k)_{f,k\in \mathbb{Z}_v}$ is defined in the following way
		\begin{equation}\label{eq-cons-1}
			\mathbf{P}(f,k)=\begin{cases}
				(f+k,i),& \mbox{if\ } k-f \in \mathcal{D}_i,\ \exists\ i\in[b];\\
				\ \ \ \ \ *, &\mbox{otherwise}.
			\end{cases}
		\end{equation} \hfill $\square$
	\end{construction}
	
	Since $\mathfrak{D}$ contains $b$ disjoint blocks $\mathcal{D}_i$ of size $g$, each column contains exactly $bg$ integer entries and $v-bg$ star entries. So the condition C$1$ and C$2$ hold. Moreover, for each integer entry $(c,i)$ and each $d \in \mathcal{D}_i$, we have
	\begin{equation}\label{eq-cons-1-1}
		k - f = d, \quad k + f = c.
	\end{equation}
	Condition C$3$-a is satisfied because \eqref{eq-cons-1-1} has a unique solution $(f,k)$, ensuring that each integer occurs at least once. Finally, if two distinct entries $\mathbf{P}(f_1,k_1) = \mathbf{P}(f_2,k_2) = (c,i)$, then the half-sum of the corresponding block elements does not belong to any block, and hence the entries $\mathbf{P}(f_1,k_1)$ and $\mathbf{P}(f_2,k_2)$ must be stars, i.e., condition C$3$-b holds. Therefore, the constructed array is a PDA. Thus, we can obtain the following result. 
	\begin{lemma}[\cite{CWWC}]\rm\label{th-PDA-NHSDP}
		Given a $(v,g,b)$ NHSDP, there exists a $(v,v,v-bg,bv)$ PDA.
	\end{lemma}  
	Let us take the $(5,2,2)$ NHSDP in Example~\ref{exam-def-nh} as an example to further illustrate Construction~\ref{cons-PDA-NHSDP}.  By Construction~\ref{cons-PDA-NHSDP}, the following array can be obtained,
	\begin{align}\label{eq-ex-P1}
		\addtocounter{MaxMatrixCols}{10}
		\setlength{\arraycolsep}{1.0pt}
		\mathbf{P}={\small
			\begin{pmatrix}
				*&(1,1)&(2,2)&(3,2)&(4,1)\\
				(1,1)&*&(3,1)&(4,2)&(0,2)\\
				(2,2)&(3,1)&*&(0,1)&(1,2)\\
				(3,2)&(4,2)&(0,1)&*&(2,1)\\
				(4,1)&(0,2)&(1,2)&(2,1)&*\\
		\end{pmatrix}}.
	\end{align}
	Clearly, $\mathbf{P}$ is a $(15,15,7,30)$ PDA. 
	
	By Lemma~\ref{th-PDA-NHSDP}, in order to obtain a PDA with linear subpacketization the authors in \cite{CWWC} introduce a construction of NHSDPs. The idea is to embed integers into a high-dimensional geometric space. Specifically, a subset $\mathcal{X}$ of $\mathbb{Z}_v$ is designed to ensure that each element in $\mathfrak{D}$ can be uniquely represented as a linear combination of elements in $\mathcal{X}$. Moreover, the half-sum of any two distinct elements within the same block can also be represented by $\mathcal{X}$, and the sets of coefficients for these two representations must be disjoint.\hfill $\square$

	Inspired by NHSDP, we propose the following novel combinatorial structure to construct $L$-continuous PDAs with linear subpacketization.
	\section{Cyclic Multi-access Non-half-sum disjoint packing}\label{sec-CMA-NHSDP}
	
	In this section, we introduce a novel combinatorial structure called cyclic multi-access non-half-sum disjoint packing (CMA-NHSDP), and further construct an $L$-continuous $(K,F,Z,S)$ PDA with linear subpacketization via the CMA-NHSDP framework.

	The authors in \cite{CWWC} first introduced NHSDP combinatorial structure to construct PDAs with subpacketization $F=K$, which is realized by constructing a Latin square and then putting stars in some integer entries. However, the PDAs constructed by the NHSDP framework cannot guarantee that the condition C$4$ of $\mathbf{Q}$ must hold. To ensure the array satisfies condition C$4$, our structure will add $L$-continuity constraint to integer entries that are replaced by stars in NHSDP.
	\begin{definition}[Cyclic Multi-access Non-half-sum disjoint packing, CMA-NHSDP]\label{def-CMA-NHSDP}\rm
		Let $\mathscr{D} \triangleq \bigcup_{i=1}^{b} \mathcal{D}_i$. For a positive integer $L$, a $(v, g, b)$ NHSDP is called an $(L, v, g, b)$ CMA-NHSDP if it satisfies:
		\begin{itemize}
			\item \textbf{$L$-continuity:}  
			The set $\mathbb{Z}_v \setminus \mathscr{D}$ can be partitioned into disjoint subsets, each consisting of exactly $L$ cyclic consecutive integers.
		\end{itemize}
		\hfill $\square$
	\end{definition}
	
	Let us take the following example with $v=7$ to further explain the concept of CMA-NHSDP and Construction~\ref{cons-PDA-NHSDP}. 
	
	\begin{example}\rm
		\label{exam-v-NHSDP}Consider $L=3$, a pair $(\mathbb{Z}_{27}, \mathfrak{D})$ with 
		\begin{align*}\label{eq-ex-d1}
			\mathfrak{D}= \{\mathcal{D}_1=&\{11, -7, 7, -11\}=\{-16,20,-20,16\},\\ \mathcal{D}_2=&\{12, -6, 6, -12\}=\{-15,21,-21,15\},\\ \mathcal{D}_3=&\{13, -5, 5, -13\}=\{-14,22,-22,14\}\}.
		\end{align*}Clearly $\mathcal{D}_1\cap\mathcal{D}_2\cap\mathcal{D}_3=\emptyset$, i.e., the first condition of Definition~\ref{def-NHSDP} holds. Moreover, $\mathbb{Z}_{27} \setminus \mathscr{D}$ can be partitioned into groups $\{26, 0, 1\}$, $\{2, 3, 4\}$, $\{ 8, 9, 10\}$, $\{17, 18, 19\}$ and $\{23, 24, 25\}$, which consists of $L = 3$ cyclic consecutive integers, thereby satisfying the condition $L$-continuity in Definition~\ref{def-CMA-NHSDP}. Now let us consider the second condition in Definition~\ref{def-NHSDP}. 
		The half-sums of any two different elements in each block in $\mathfrak{D}$ are as follows, 
		\begin{equation}\label{eq:ex D1 D2}
			\begin{aligned}
				& {\rm in}\  \mathcal{D}_1:  \frac{-7+7}{2}=0,\  
				\frac{-7-11}{2}=-9,\ 
				\frac{-7+11}{2}=2,\  
				\frac{7-11}{2}=-2, \ 
				\frac{7+11}{2}=9,\  		\frac{-11+11}{2}=0,\\
				&{\rm in}\  \mathcal{D}_2: \frac{-6+6}{2}=0,\  
				\frac{-6-12}{2}=-9,\ 
				\frac{-6+12}{2}=9,\  
				\frac{6-12}{2}=-3, \ 
				\frac{6+12}{2}=3,\  
				\frac{-12+12}{2}=0,\\
				& {\rm in}\  \mathcal{D}_3:  \frac{-5+5}{2}=0,\  
				\frac{-5-13}{2}=-9,\ 
				\frac{-5+13}{2}=4,\  
				\frac{5-13}{2}=-4, \ 
				\frac{5+13}{2}=9,\  		\frac{-13+13}{2}=0,\\
			\end{aligned}
		\end{equation}We have $\{0,\pm 2,\pm 3,\pm 4 \pm 9\}\cap\mathcal{D}_1=\{0,\pm 2,\pm 3,\pm 4 \pm 9\}\cap\mathcal{D}_2=\{0,\pm 2,\pm 3,\pm 4 \pm 9\}\cap\mathcal{D}_3=\emptyset$, i.e., the second condition of Definition~\ref{def-NHSDP} holds. So $(\mathbb{Z}_{27},\mathfrak{D})$ is a $(3,27,4,3)$ CMA-NHSDP.

		Based on this CMA-NHSDP, the following array $\mathbf{P}$ can be obtained by Construction~\ref{cons-PDA-NHSDP}. Due to the large number of columns in $\mathbf{P}$, we will use $f+k\!,\! i$ instead of $(f+k,i)$.
		\begin{align}\label{eq-ex-P1}
			\addtocounter{MaxMatrixCols}{27}
			\setlength{\arraycolsep}{3pt}
			\mathbf{\tiny 
				\begin{pmatrix}
					* & * & * & * & * &5\!,\! 3 &6\!,\! 2 &7\!,\! 1 & * & * & * &11\!,\! 1 &12\!,\! 2 &13\!,\! 3 &14\!,\! 3 &15\!,\! 2 &16\!,\! 1 & * & * & * &20\!,\! 1 &21\!,\! 2 &22\!,\! 3 & * & * & * & * \\
					* & * & * & * & * & * &7\!,\! 3 &8\!,\! 2 &9\!,\! 1 & * & * & * &13\!,\! 1 &14\!,\! 2 &15\!,\! 3 &16\!,\! 3 &17\!,\! 2 &18\!,\! 1 & * & * & * &22\!,\! 1 &23\!,\! 2 &24\!,\! 3 & * & * & * \\
					* & * & * & * & * & * & * &9\!,\! 3 &10\!,\! 2 &11\!,\! 1 & * & * & * &15\!,\! 1 &16\!,\! 2 &17\!,\! 3 &18\!,\! 3 &19\!,\! 2 &20\!,\! 1 & * & * & * &24\!,\! 1 &25\!,\! 2 &26\!,\! 3 & * & * \\
					* & * & * & * & * & * & * & * &11\!,\! 3 &12\!,\! 2 &13\!,\! 1 & * & * & * &17\!,\! 1 &18\!,\! 2 &19\!,\! 3 &20\!,\! 3 &21\!,\! 2 &22\!,\! 1 & * & * & * &26\!,\! 1 &0\!,\! 2 &1\!,\! 3 & * \\
					* & * & * & * & * & * & * & * & * &13\!,\! 3 &14\!,\! 2 &15\!,\! 1 & * & * & * &19\!,\! 1 &20\!,\! 2 &21\!,\! 3 &22\!,\! 3 &23\!,\! 2 &24\!,\! 1 & * & * & * &1\!,\! 1 &2\!,\! 2 &3\!,\! 3 \\
					5\!,\! 3 & * & * & * & * & * & * & * & * & * &15\!,\! 3 &16\!,\! 2 &17\!,\! 1 & * & * & * &21\!,\! 1 &22\!,\! 2 &23\!,\! 3 &24\!,\! 3 &25\!,\! 2 &26\!,\! 1 & * & * & * &3\!,\! 1 &4\!,\! 2 \\
					6\!,\! 2 &7\!,\! 3 & * & * & * & * & * & * & * & * & * &17\!,\! 3 &18\!,\! 2 &19\!,\! 1 & * & * & * &23\!,\! 1 &24\!,\! 2 &25\!,\! 3 &26\!,\! 3 &0\!,\! 2 &1\!,\! 1 & * & * & * &5\!,\! 1 \\
					7\!,\! 1 &8\!,\! 2 &9\!,\! 3 & * & * & * & * & * & * & * & * & * &19\!,\! 3 &20\!,\! 2 &21\!,\! 1 & * & * & * &25\!,\! 1 &26\!,\! 2 &0\!,\! 3 &1\!,\! 3 &2\!,\! 2 &3\!,\! 1 & * & * & * \\
					* &9\!,\! 1 &10\!,\! 2 &11\!,\! 3 & * & * & * & * & * & * & * & * & * &21\!,\! 3 &22\!,\! 2 &23\!,\! 1 & * & * & * &0\!,\! 1 &1\!,\! 2 &2\!,\! 3 &3\!,\! 3 &4\!,\! 2 &5\!,\! 1 & * & * \\
					* & * &11\!,\! 1 &12\!,\! 2 &13\!,\! 3 & * & * & * & * & * & * & * & * & * &23\!,\! 3 &24\!,\! 2 &25\!,\! 1 & * & * & * &2\!,\! 1 &3\!,\! 2 &4\!,\! 3 &5\!,\! 3 &6\!,\! 2 &7\!,\! 1 & * \\
					* & * & * &13\!,\! 1 &14\!,\! 2 &15\!,\! 3 & * & * & * & * & * & * & * & * & * &25\!,\! 3 &26\!,\! 2 &0\!,\! 1 & * & * & * &4\!,\! 1 &5\!,\! 2 &6\!,\! 3 &7\!,\! 3 &8\!,\! 2 &9\!,\! 1 \\
					11\!,\! 1 & * & * & * &15\!,\! 1 &16\!,\! 2 &17\!,\! 3 & * & * & * & * & * & * & * & * & * &0\!,\! 3 &1\!,\! 2 &2\!,\! 1 & * & * & * &6\!,\! 1 &7\!,\! 2 &8\!,\! 3 &9\!,\! 3 &10\!,\! 2 \\
					12\!,\! 2 &13\!,\! 1 & * & * & * &17\!,\! 1 &18\!,\! 2 &19\!,\! 3 & * & * & * & * & * & * & * & * & * &2\!,\! 3 &3\!,\! 2 &4\!,\! 1 & * & * & * &8\!,\! 1 &9\!,\! 2 &10\!,\! 3 &11\!,\! 3 \\
					13\!,\! 3 &14\!,\! 2 &15\!,\! 1 & * & * & * &19\!,\! 1 &20\!,\! 2 &21\!,\! 3 & * & * & * & * & * & * & * & * & * &4\!,\! 3 &5\!,\! 2 &6\!,\! 1 & * & * & * &10\!,\! 1 &11\!,\! 2 &12\!,\! 3 \\
					14\!,\! 3 &15\!,\! 3 &16\!,\! 2 &17\!,\! 1 & * & * & * &21\!,\! 1 &22\!,\! 2 &23\!,\! 3 & * & * & * & * & * & * & * & * & * &6\!,\! 3 &7\!,\! 2 &8\!,\! 1 & * & * & * &12\!,\! 1 &13\!,\! 2 \\
					15\!,\! 2 &16\!,\! 3 &17\!,\! 3 &18\!,\! 2 &19\!,\! 1 & * & * & * &23\!,\! 1 &24\!,\! 2 &25\!,\! 3 & * & * & * & * & * & * & * & * & * &8\!,\! 3 &9\!,\! 2 &10\!,\! 1 & * & * & * &14\!,\! 1 \\
					16\!,\! 1 &17\!,\! 2 &18\!,\! 3 &19\!,\! 3 &20\!,\! 2 &21\!,\! 1 & * & * & * &25\!,\! 1 &26\!,\! 2 &0\!,\! 3 & * & * & * & * & * & * & * & * & * &10\!,\! 3 &11\!,\! 2 &12\!,\! 1 & * & * & * \\
					* &18\!,\! 1 &19\!,\! 2 &20\!,\! 3 &21\!,\! 3 &22\!,\! 2 &23\!,\! 1 & * & * & * &0\!,\! 1 &1\!,\! 2 &2\!,\! 3 & * & * & * & * & * & * & * & * & * &12\!,\! 3 &13\!,\! 2 &14\!,\! 1 & * & * \\
					* & * &20\!,\! 1 &21\!,\! 2 &22\!,\! 3 &23\!,\! 3 &24\!,\! 2 &25\!,\! 1 & * & * & * &2\!,\! 1 &3\!,\! 2 &4\!,\! 3 & * & * & * & * & * & * & * & * & * &14\!,\! 3 &15\!,\! 2 &16\!,\! 1 & * \\
					* & * & * &22\!,\! 1 &23\!,\! 2 &24\!,\! 3 &25\!,\! 3 &26\!,\! 2 &0\!,\! 1 & * & * & * &4\!,\! 1 &5\!,\! 2 &6\!,\! 3 & * & * & * & * & * & * & * & * & * &16\!,\! 3 &17\!,\! 2 &18\!,\! 1 \\
					20\!,\! 1 & * & * & * &24\!,\! 1 &25\!,\! 2 &26\!,\! 3 &0\!,\! 3 &1\!,\! 2 &2\!,\! 1 & * & * & * &6\!,\! 1 &7\!,\! 2 &8\!,\! 3 & * & * & * & * & * & * & * & * & * &18\!,\! 3 &19\!,\! 2 \\
					21\!,\! 2 &22\!,\! 1 & * & * & * &26\!,\! 1 &0\!,\! 2 &1\!,\! 3 &2\!,\! 3 &3\!,\! 2 &4\!,\! 1 & * & * & * &8\!,\! 1 &9\!,\! 2 &10\!,\! 3 & * & * & * & * & * & * & * & * & * &20\!,\! 3 \\
					22\!,\! 3 &23\!,\! 2 &24\!,\! 1 & * & * & * &1\!,\! 1 &2\!,\! 2 &3\!,\! 3 &4\!,\! 3 &5\!,\! 2 &6\!,\! 1 & * & * & * &10\!,\! 1 &11\!,\! 2 &12\!,\! 3 & * & * & * & * & * & * & * & * & * \\
					* &24\!,\! 3 &25\!,\! 2 &26\!,\! 1 & * & * & * &3\!,\! 1 &4\!,\! 2 &5\!,\! 3 &6\!,\! 3 &7\!,\! 2 &8\!,\! 1 & * & * & * &12\!,\! 1 &13\!,\! 2 &14\!,\! 3 & * & * & * & * & * & * & * & * \\
					* & * &26\!,\! 3 &0\!,\! 2 &1\!,\! 1 & * & * & * &5\!,\! 1 &6\!,\! 2 &7\!,\! 3 &8\!,\! 3 &9\!,\! 2 &10\!,\! 1 & * & * & * &14\!,\! 1 &15\!,\! 2 &16\!,\! 3 & * & * & * & * & * & * & * \\
					* & * & * &1\!,\! 3 &2\!,\! 2 &3\!,\! 1 & * & * & * &7\!,\! 1 &8\!,\! 2 &9\!,\! 3 &10\!,\! 3 &11\!,\! 2 &12\!,\! 1 & * & * & * &16\!,\! 1 &17\!,\! 2 &18\!,\! 3 & * & * & * & * & * & * \\
					* & * & * & * &3\!,\! 3 &4\!,\! 2 &5\!,\! 1 & * & * & * &9\!,\! 1 &10\!,\! 2 &11\!,\! 3 &12\!,\! 3 &13\!,\! 2 &14\!,\! 1 & * & * & * &18\!,\! 1 &19\!,\! 2 &20\!,\! 3 & * & * & * & * & * \\
			\end{pmatrix}}.
		\end{align}
		We can find that the stars occur exactly $Z=v-bg=27-3\times4=15$ times in each column of $\mathbf{P}$, and appear in each row as $L=3$ cyclic consecutive. So the conditions C$2$ and C$4$ hold. In addition, there are exactly $S=81$ vectors in  $\mathbf{P}$, and each vector occurs at most once in each row and each column. i.e., the conditions C$2$ and C$3$-a of Definition~\ref{def-PDA} hold. Finally, let us consider condition C$3$-b. Let us first consider the entries $\mathbf{P}(5,0)=\mathbf{P}0,5)=(5,3)$. We have $\mathbf{P}(5,5)=\mathbf{P}(0,0)=*$ which satisfies condition C$3$-b. Similarly, we can check that all the vectors in $\mathbf{P}$ satisfy condition C$3$-b) of Definition~\ref{def-PDA}. Therefore, $\mathbf{P}$ is an $L$-continuous $(27,27,15,81)$ PDA that can realize a MACC scheme with memory ratio $\frac{M}{N}=\frac{5}{27}$, subpacketization $F=K=27$, and load $R=3$.

		Now, let us consider the existing schemes with $K=27$ and $M/N=5/27$.
		\begin{itemize}
			\item The WCWL scheme: In \cite{WCWL} we have a $(K=27,L=3,M,N)$ WCWL scheme with the memory ratio $M/N = 5/27$, the subpacketization $F_{\text{WCWL}} = 2\lfloor\frac{27}{27-15+1}\rfloor 7 = 28 > 27=F$, the transmission load $R_{\text{WCWL}} = \frac{27-15}{2\lfloor\frac{27}{27-15+1}\rfloor} =3= 3 = R$. We can see that our scheme has lower subpacketization while maintaining the same transmission load and memory ratio;
			\item  The CW scheme: In \cite{CWZW} we have a $(K=27,L=3,M,N)$ CW scheme with the memory ratio $M/N=5/27$, the subpacketization $F_{\text{CW}} =167076 >27= F$, the transmission load $R_{\text{CW}} = 2 = R$. Clearly our scheme achieves a significantly reduced subpacketization, while maintaining the same subpacketization, transmission load and memory ratio.
		\end{itemize}
		\hfill $\square$
	\end{example}
	In fact, for any parameters $L$, $v$, $g$ and $b$, if there is exists an $(L,v,g,b)$ CMA-NHSDP, we can also obtain an $L$-continuous PDA with the following theorem by Construction~\ref{cons-PDA-NHSDP}. That is the following result. 
	\begin{theorem}[$L$-continuous PDA via CMA-NHSDP]\rm\label{th-PDA via CMA}
		Given an $(L,v,g,b)$ CMA-NHSDP, we can obtain an $L$-continuous $(v,v,v-bg,bv)$ PDA which generates a $(K=v,L,M,N)$ MACC scheme with memory ratio $\frac{M}{N}=\frac{v-bg}{vL}$, coded caching gain $g$, subpacketization $F=v$, and  transmission load $R=b$.\hfill $\square$
	\end{theorem} 
	The reason is as follows.
	\begin{remark}\rm
		By lemma~\ref{th-PDA-NHSDP}, the array constructed by CMA-NHSDP is a PDA since it is the special case of NHSDP. From Construction~\ref{cons-PDA-NHSDP}, we have $\mathbf{P}(f,k) = *$ when $k-f \in \mathbb{Z}_v \setminus \mathscr{D}$, implying that the set of columns containing the symbol `*' is $\{k \mid k-f \in \mathbb{Z}_v \setminus \mathscr{D}\} = f + (\mathbb{Z}_v \setminus \mathscr{D})$ for a fixed row $f$. The $L$-continuity condition guarantees that $\mathbb{Z}_v \setminus \mathscr{D}$ can be partitioned into disjoint subsets of exactly $L$ cyclic consecutive integers. So the symbol `*' in each row $f$ must appear as cyclic consecutive runs of length $L$, i.e., condition C$4$ in Definition~\ref{def-PDA} holds. 
	\end{remark}

	\begin{remark}\rm
		\label{re-even-case}  When $K$ is even, we can add a virtual user into the MACC system, making the efficient number of users $K+1$ which can be solved by the  $(L,K+1,g,b)$ CMA-NHSDP in Theorem~\ref{th-PDA via CMA}. 
	\end{remark}
	
	By Theorem \ref{th-PDA via CMA}, we can obtain an $L$-continuous PDA by constructing a CMA-NHSDP. Thereby it is meaningful to construct CMA-NHSDP with good performance.

	\section{Constructions of PDAs via CMA-NHSDP}
	\label{sec-Construct-CMA-NHSDP}
	
	In this section, we propose a novel construction framework for CMA-NHSDPs. Moreover, we derive the good solution by transforming parameter selection into an integer optimization problem  under this framework. 
	
	Our construction is based on the NHSDPs construction framework to obtain an initial subset $\mathcal{X}$ of $\mathbb{Z}_v$ and $\mathfrak{D}$, thereby ensuring the conditions in Definition~\ref{def-NHSDP} hold. Since the $L$-continuity condition in Definition~\ref{def-CMA-NHSDP} also needs to be guaranteed, we will introduce an offset function to directly adjust the subsets $\mathcal{X}$ and $\mathfrak{D}$. Such that after sorting the elements in resulting $\mathscr{D}$, the difference between any two neighboring elements must be $kL+1$, where $k\in\mathbb{N}$. By finding an appropriate  offset function and subset $\mathcal{X}$, we obtain the following main construction in this paper.\\
	
	
	\begin{construction}\rm
		\label{constr-2}
		For any positive odd integer $L$  and $n$ positive integers $m_1$, $m_2$, $\ldots$, $m_n$, let $\mathcal{A}:=[m_1]\times [m_2]\times \cdots\times[m_n]$. Define the following recursive functions:
		\begin{equation}
			\label{eq-RF}
			f(i)=\begin{cases}
				m_1\ \ \ \ \ \ \ \ \ \ \ \ \ \ \ \ \ \ \ \ \ \ \ \ &\ \text{if}\ i=1,\\
				m_i \left( 2 \sum_{j=1}^{i-1} \left( f(j) + g(j) \right) + 1 \right) &\ \text{if}\ i \geq 2. 
			\end{cases}
		\end{equation}
		\begin{equation}
			\label{eq-RF-g}
			g(i)=\begin{cases}
				\frac{L-1}{2}\ \ \ \ \ \ \ \ \ \ \ \ \ \ \ \ \ \ \ \ \ \ &\ \text{if}\ i=1,\\
				\left\langle -f(i-1) \right\rangle_L &\ \text{if}\ i \geq 2. 
			\end{cases}
		\end{equation}
		and $\mathcal{X}:=\left\{x_i=\frac{f(i)}{m_i} \;\big|\; i\in [n]\right\}$. We can construct a family
		$\mathfrak{D}=\{\mathcal{D}_{\bf a} \mid {\bf a}=(a_1,a_2,\ldots,a_n)\in \mathcal{A}\}$ where $\mathcal{D}_{\bf a}$ is defined as  
		\begin{equation}
			\label{eq-family}
			\mathcal{D}_{\bf a} = \left\{\alpha_1 (a_1 x_1 + g(1)) + \alpha_2 (a_2 x_2 + g(2)) + \cdots + \alpha_n (a_n x_n + g(n)) \;\Big|\; \alpha_i\in\{-1,1\}, i\in[n]\right\},
		\end{equation}
		for each vector ${\bf a}=(a_1,a_2,\ldots,a_n)\in \mathcal{A}$. By the above construction, $\mathfrak{D}$ has $m_1 \times m_2 \times \cdots \times m_n$ blocks, and each block $\mathcal{D}_{\bf a}$ has $2^n$ integers. 
		\hfill $\square$ \\
		%
	\end{construction}

	
	Let us take the following example to illustrate Construction~\ref{constr-2}.
	
	\begin{example}\rm \label{exam-n=3}
		When $L=3$, $n=3$, $m_1=3$, $m_2=1$, $m_3=1$, we have $\mathcal{A}=[3] \times [1] \times [1]$. From \eqref{eq-RF} and \eqref{eq-RF-g}, when $i=1$,$2$ and $3$ we have 
		\begin{align*}
			f(1) &= m_1 = 3, \quad g(1) = 
			\frac{L-1}{2}= \frac{3-1}{2}  = 1;\\
			f(2)= m_2(2(f(1) + g(1))& + 1) = 1(2(3 + 1) + 1) = 9,\quad g(2)= \left\langle -f(i-1) \right\rangle_L = \left\langle -3 \right\rangle_3 = 0;\\f(3) = m_3(2(f(1) + g(1))+2(f(2) + g(2))& + 1) = 1(2(3 + 1)+2(9 + 0) + 1) = 27, \quad g(3) = \left\langle -f(i-1) \right\rangle_L = \left\langle -9 \right\rangle_3 = 0;
		\end{align*}
		respectively. Then we have the subset 
		\begin{align*}
			\mathcal{X} &= \left\{x_1=\frac{f_1}{m_1}=\frac{3}{3}=1, x_2=\frac{f(2)}{m_2}=\frac{9}{1}=9,x_3=\frac{f(2)}{m_3}=\frac{27}{1}=27\right\}.
		\end{align*}
		Now let us consider $\mathfrak{D}$ in Construction~\ref{constr-2}. When ${\bf a}=(1,1,1)$ and $(x_1,x_2,x_3)=(1,9,27)$, from \eqref{eq-family}, then we have 
		\begin{align*}
			\mathcal{D}_{(1,1,1)} &=\left\{\alpha_1 (a_1 x_1 + g(1)) + \alpha_2 (a_2 x_2 + g(2))+ \alpha_2 (a_3 x_3 + g(3))  \;\Big|\; \alpha_1,\alpha_2,\alpha_3\in\{-1,1\}\right\} \\
			&= \{2 \cdot \alpha_1+9 \cdot \alpha_2+27 \cdot \alpha_3 \mid \alpha_1,\alpha_2,\alpha_3\in\{-1,1\} \} \\
			&= \{38, -16, 20, -34, 34, -20, 16, -38\}.
		\end{align*}
		\  For instance, the integer $38$ in $\mathcal{D}_{(1,1,1)}$ can be obtained by $2+9+27=38$. The number of all possible $(\alpha_1, \alpha_2, \alpha_3)$ is $2^3=8$; thus the block $\mathcal{D}_{(1,1,1)}$ contains $8$ different integers. Similarly, we can obtain the following blocks.

		\begin{equation}\label{eq-points}
			\begin{split}
				\mathfrak{D}=\{&\mathcal{D}_{(1,1,1)}=\{38, -16, 20, -34, 34, -20, 16, -38
				\},\\
				&\mathcal{D}_{(2,1,1)}=\{39, -15, 21, -33, 33, -21, 15, -39
				\},\\ 
				&\mathcal{D}_{(3,1,1)}=\{40, -14, 22, -32, 32, -22, 14, -40
				\}\}.
			\end{split}
		\end{equation}
		By the above equation, $\mathfrak{D}$ has $m_1m_2m_3=3$ blocks and the intersection of any blocks is empty. The minimum and  maximum values of $\mathfrak{D}$ are $-40$ and $40$, respectively. To ensure that these two sets $[40:0]$ and $[1:40]$ do not have overlap in $\mathbb{Z}_v$, the value of $v$ must satisfy $v \geq 2\times 40+1=81$, since $0$ is also in $\mathbb{Z}_{v}$. Next, let us verify that $(\mathbb{Z}_{81},\mathfrak{D})$ is a $(3,81,8,3)$ CMA-NHSDP. Now let we consider the half-sum of $38$ and $16$, i.e., $\frac{38+16}{2}=27$, which is not in any block of $\mathfrak{D}$. Similarly, we can check the half-sum of any two different integers in each block of $\mathfrak{D}$. Finally, we can check that the sorted set $\mathscr{D} = \{\pm 14, \pm 15, \pm 16, \pm 20, \pm 21, \pm 22, \pm 32, \pm 33, \pm 34, \pm 38, \pm 39, \pm 40\}$ satisfies that its complement in $\mathbb{Z}_7$ can be partitioned into groups of $3$ cyclic consecutive integers, such as $\{17, 18, 19\}$.
	\end{example}
	\hfill $\square$

	By Theorem~\ref{th-PDA via CMA} we have a $(81,81,57,243)$ PDA which generates a $81$-division $(81,3,M,N)$ MACC scheme with $M/N=\frac{19}{81}$, coded caching gain $g=8$, and the transmission load $b=3$. Now let us consider the existing schemes with $K=81$.
	\begin{itemize}
		\item  The WCWL scheme (see Table~\ref{tab-knowschemes}): When $K=81$ and $t=19$ in \cite{WCWL} we have a WCWL scheme with $K=81$, the subpacketization $F=81$, and the transmission load $\frac{K-tL}{2\lfloor \frac{K}{K-tL+1}\rfloor}=\frac{24}{6}=4$. Clearly our scheme has a lower transmission load while maintaining the same memory ratio and subpacketization;
		\item The CW scheme (see Table~\ref{tab-knowschemes}): An exhaustive computer search shows that when $t=19$, we can obtain a CW scheme with $M/N=\frac{19}{81}$, an appropriate subpacketization $4.44\times10^{13}$, and the transmission load $R=\frac{K - tL}{t+1}=6>3$. It is evident that at the same memory ratio, the subpacketization and transmission load is larger than our proposed scheme.
	\end{itemize}
	
	\begin{remark}\rm
		\label{re-even-case}  When $L$ is even, we take the $(L-1, v, g, b)$ CMA-NHSDP scheme, and retrieve one additional neighboring node for each $L$ cyclic consecutive cache nodes retrieved by the user. 
	\end{remark} 
	
	By Construction~\ref{constr-2}, for any  $n$ and $m_1$, $m_2$, $\ldots$, $m_n$ are positive integers, suppose the vector $\mathbf{a}=(a_1,a_2,\ldots,a_n)=(m_1,m_2,\ldots,m_n) \in \mathcal{A}$. The corresponding block  $\mathcal{D}_{\mathbf{a}}$ is given by  $$\mathcal{D}_{\bf a} = \left\{\alpha_1 (f(1) + g(1)) + \alpha_2 (f(2) + g(2)) + \cdots + \alpha_n (f(n) + g(n)) \;\Big|\; \alpha_i\in\{-1,1\}, i\in[n]\right\},$$
	which contains the integers   $-(f(1) + g(1))-(f(2) + g(2))-\dots-(f(n) + g(n))$ and $(f(1) + g(1))+(f(2) + g(2))+\dots+(f(n) + g(n))$. To ensure that  $[-(f(1) + g(1))-(f(2) + g(2))-\dots-(f(n) + g(n)):0]$ and $[1:(f(1) + g(1))+(f(2) + g(2))+\dots+(f(n) + g(n))]$ can not overlap in $\mathbb{Z}_{v}$, the value of $v$ must satisfy $$v\geq 2((f(1) + g(1))+(f(2) + g(2))+\dots+(f(n) + g(n)))+1.$$ 
	For any positive integers $n$ and $m_1, \cdots, m_n$, we define that $ \phi(m_1,m_2,\ldots,m_n):=\sum_{i=1}^{n}\left( f(i) + g(i) \right)$
	\begin{equation} 
		\begin{aligned}
			\label{eq-sum}
			2\phi(m_1,m_2,\ldots,m_n)+1=&2(1+2m_n)\sum_{i=1}^{n-1}(f(i) + g(i) )+2m_n+2g(n)+1\\
			=&(1+2m_n)\left(2\sum_{i=1}^{n-1}(f(i) + g(i)) +1\right)+2g(n)\\
			=&(1+2m_n)\left((1+2m_{n-1})(2\sum_{i=1}^{n-2}(f(i) + g(i) )+1)+2g(n-1)\right)+2g(n)\\
			=&\prod_{i=1}^{n}(1+2m_i)+2\sum_{i=1}^{n}\left(g(i)\prod_{j=i+1}^{n}(1+2m_j)\right).
		\end{aligned}
	\end{equation}
	
	\begin{theorem}\rm
		\label{th-main-2}
		For any $n$ and $m_1$, $m_2$, $\ldots$, $m_n$  are positive integers and for any odd positive integer $v\geq 2\phi(m_1,m_2,\ldots,m_n)+1$ defined in \eqref{eq-sum}, the pair $(\mathbb{Z}_{v},\mathfrak{D})$ generated in Construction~\ref{constr-2} is an $(L,v,2^n,\prod_{i=1}^{n}m_i)$ CMA-NHSDP.
		\hfill $\square$ 
	\end{theorem}
	
	By using the above CMA-NHSDP construction, we can obtain the following MACC schemes.

	\begin{theorem}\rm
		\label{th-main-PDA}
		Given an $(L,v,2^n,\prod_{i=1}^{n}m_i)$ CMA-NHSDP where $v\geq 2\phi(m_1,m_2,\ldots,m_n)+1$ defined in \eqref{eq-sum}, there exist a $F$-division $(K=v,L,M,N)$ MACC scheme with memory ratio $M/N=\frac{v-2^n\prod_{i=1}^{n}m_i}{vL}$, subpacketization $F=v$, and transmission load $R=\prod_{i=1}^{n}m_i$. 
		\hfill $\square$ 
	\end{theorem}
	Next, given the parameters $v$ and $n$ (recall that $K=v$ and $g=2^n$), we consider how to select $m_1,\ldots,m_n$ such that the memory ratio is minimized. That is, we aim to find the MACC scheme with the minimum required memory ratio while fixing the coded caching gain. Considering the constraint $v\geq 2\phi(m_1,m_2,\ldots,m_n)+1$ into the optimization problem for the selection of $m_1,m_2,\ldots,m_n$, we have 
	\begin{equation}\label{eq-good}
		\begin{aligned}
			\text{Problem 1.}  \ \ \  & \textbf{Maximize } \text{function } f=\prod_{i=1}^{n}m_i \\
			&\textbf{Constrains: } m_1,\ldots,m_n \in \mathbb{Z}^{+},\\
			&\quad\quad\quad \quad \quad  \prod_{i=1}^{n}(1+2m_i)+2\sum_{i=1}^{n}\left(g(i)\prod_{j=i+1}^{n}(1+2m_j)\right) \leq v.
		\end{aligned}
	\end{equation}
	Since this equation \eqref{eq-good} involves modulo arithmetic, it can not be solved directly. However, through abundant computer calculations, we have found that a good solution can be obtained when $m_i$ satisfies the following relationship.
	
	%

\begin{figure}[ht]
		\centering
		\includegraphics[height=6cm]{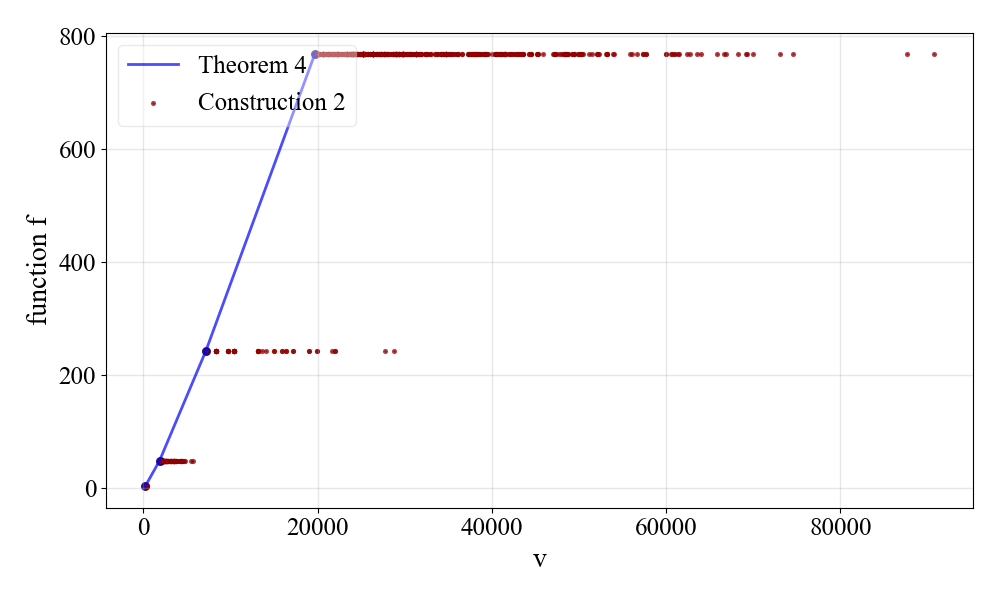}
		\vskip 0.2cm
		\caption{Tradeoff between $v$ and $f$}\label{good_Compare}
	\end{figure}

	\begin{theorem}\rm
		\label{th:Lagrange}
		A good solution (with closed-form) to Problem~1 is 
		\begin{align}
			\label{eq:m=m2}
			m_1=Lm_2,\ \  m_2=m_3=m_4=\cdots=m_n=\left\lfloor \frac{(\frac{v}{L})^{1/n}-1}{2} \right\rfloor=\left\lfloor\frac{q-1}{2}\right\rfloor. 
		\end{align}
		Under the selection in~\eqref{eq:m=m2}, the resulting $L$-continuous PDA has the parameters 
		\begin{equation*}
			K=q^nL,\ \ F=q^nL,\ \ Z=q^nL-{\left(2\lfloor{\frac{q-1}{2}}\rfloor\right)}^{n}L,\ \ S=\lfloor\frac{q-1}{2}\rfloor^{n}q^nL^2,
		\end{equation*} which can realize a MACC scheme with the memory ratio $\frac{M}{N}=\frac{1}{L}-\frac{{\left(2\lfloor{\frac{q-1}{2}}\rfloor\right)}^{n}}{q^nL}$ and the transmission load $R=\lfloor\frac{q-1}{2}\rfloor^{n}L$.
		
		\hfill $\square$ 
	\end{theorem}
	For any positive odd integer $q \geq 3$, the solution in~\eqref{eq:m=m2} is a good solution to Problem~1. Then the following result can be obtained.
	\begin{remark}[Good solution of Problem 1]\rm
		\label{re-optimal}
		When $q \geq 3$ and $L$ in Theorem~\ref{th:Lagrange} is an odd integer, we have an $L$-continuous $(K=q^nL, F=q^nL, Z=q^nL-(q-1)^{n}L, S=(\frac{q-1}{2})^{n}q^nL^2)$ PDA which realizes a $F$-division $(K=v,L,M,N)$ MACC scheme with the memory ratio $\frac{M}{N}=\frac{1}{L}-\frac{(q-1)^{n}}{q^nL}$, subpacketization $F=q^nL$, and transmission load $R=(\frac{q-1}{2})^{n}L$.
	\end{remark}
	We now compare the performance of Theorem~\ref{th:Lagrange} and Construction~\ref{constr-2} through a concrete example. Let $n=4$, $L=3$, and let $m_2$ vary in ${1,2,3,4}$. The corresponding parameters obtained from the two approaches are summarized in Fig.~\ref{good_Compare}. It can be observed that Theorem~\ref{th:Lagrange} yields a smaller value of $v$ while maintaining the same objective function value $f$.
	\\
	\section{Performance analysis}
	\label{sec-perf-ana}
	In this section, we will present theoretical and numerical comparisons with the existing schemes \cite{RK,WCWL,HKD,CWLZC,RK2021,SR,SPE,SR_arX2021,MR,NR} respectively to show the performance of our new scheme in Theorem~\ref{th:Lagrange}.
	\subsection{Theoretical comparisons} 
	
	\subsubsection{Comparison with the RK1 scheme in \cite{RK}} 
	Let $K=q^nL$ and $t=q^n-2^n\lfloor\frac{q-1}{2}\rfloor^{n}$ in \cite{RK}, i.e., the $10$th row of Table~\ref{tab-knowschemes}. 
	We compare the RK1 scheme with the proposed scheme in the following cases:
	
	\begin{align*}
		F_{\text{RK1}}&=\frac{K}{t}\cdot \binom{K - tL+t-1}{t-1}=\frac{q^nL}{q^n-2^n\lfloor\frac{q-1}{2}\rfloor^{n}}\cdot \binom{q^n+2^n\lfloor\frac{q-1}{2}\rfloor^{n}(L-1)-1}{q^n-2^n\lfloor\frac{q-1}{2}\rfloor^{n}-1},\\ R_{\text{RK1}}&=K(1-\frac{LM}{N})^2
		=\frac{2^{2n}\lfloor\frac{q-1}{2}\rfloor^{2n}L}{q^n}.
	\end{align*}
	By Theorem~\ref{th:Lagrange}, we have the following result.
	\begin{align*}
		\frac{F_{\text{RK1}}}{F}&=\frac{\frac{q^nL}{q^n-2^n\lfloor\frac{q-1}{2}\rfloor^{n}}\cdot \binom{q^n+2^n\lfloor\frac{q-1}{2}\rfloor^{n}(L-1)-1}{q^n-2^n\lfloor\frac{q-1}{2}\rfloor^{n}-1}}{q^nL}=\frac{ \binom{q^n+2^n\lfloor\frac{q-1}{2}\rfloor^{n}(L-1)-1}{q^n-2^n\lfloor\frac{q-1}{2}\rfloor^{n}-1}}{q^n-2^n\lfloor\frac{q-1}{2}\rfloor^{n}} \geq \frac{q^n-2^n\lfloor\frac{q-1}{2}\rfloor^{n}}{q^n-2^n\lfloor\frac{q-1}{2}\rfloor^{n}}=1, \\
		\frac{R_{\text{RK1}}}{R}&=\frac{\frac{2^{2n}\lfloor\frac{q-1}{2}\rfloor^{2n}L}{q^n}}{\lfloor\frac{q-1}{2}\rfloor^{n}L}
		=\frac{2^{2n}\lfloor\frac{q-1}{2}\rfloor^{n}}{q^n}\approx(2-\frac{2}{q})^n
	\end{align*}
	As a result, compared to the RK1 scheme in \cite{RK},  our scheme either obtaining smaller or same subpacketization. The multiplicative reduction amount by our transmission is $(2-\frac{2}{q})^n$. Only when $K \geq 3^nL$ ,the formula $(2-\frac{2}{q})^n\geq1$ always holds.
	
	\subsubsection{Comparison with the CW scheme in \cite{CWLZC}} 
	Let $K=q^nL$ and $t=q^n-2^n\lfloor\frac{q-1}{2}\rfloor^{n}$ in \cite{CWLZC}, i.e., the $10$th row of Table~\ref{tab-knowschemes}. 
	We compare the CW scheme with the proposed scheme in the following cases:
	
	\begin{align*}
		F_{\text{CW}}&=K\cdot \binom{K - t(L-1)}{t}=q^nL\cdot \binom{q^n+2^n\lfloor\frac{q-1}{2}\rfloor^{n}(L-1)}{q^n-2^n\lfloor\frac{q-1}{2}\rfloor^{n}},\\ R_{\text{CW}}&=\frac{K - tL}{t+1}
		=\frac{2^n\lfloor\frac{q-1}{2}\rfloor^{n}L}{q^n-2^n\lfloor\frac{q-1}{2}\rfloor^{n}+1}.
	\end{align*}
	By Theorem~\ref{th:Lagrange}, we have the following result.
	\begin{align*}
		\frac{F_{\text{CW}}}{F}&=\frac{q^nL\cdot \binom{q^n+2^n\lfloor\frac{q-1}{2}\rfloor^{n}(L-1)}{q^n-2^n\lfloor\frac{q-1}{2}\rfloor^{n}}}{q^nL}=\binom{q^n-2^n\lfloor\frac{q-1}{2}\rfloor^{n}+2^n\lfloor\frac{q-1}{2}\rfloor^{n}L}{q^n-2^n\lfloor\frac{q-1}{2}\rfloor^{n}}, \\
		\frac{R_{\text{CW}}}{R}&=\frac{\frac{2^n\lfloor\frac{q-1}{2}\rfloor^{n}L}{q^n-2^n\lfloor\frac{q-1}{2}\rfloor^{n}+1}}{\lfloor\frac{q-1}{2}\rfloor^{n}L}=\frac{2^n}{q^n-2^n\lfloor\frac{q-1}{2}\rfloor^{n}+1}\\
	\end{align*}
	As a result, compared to the CW scheme in \cite{CWLZC},  our scheme either obtaining smaller subpacketization. The multiplicative reduction amount by our transmission is $\frac{2^n}{q^n-2^n\lfloor\frac{q-1}{2}\rfloor^{n}+1}$. Only when $K=qL$ ,the formula $\frac{2^n}{q^n-2^n\lfloor\frac{q-1}{2}\rfloor^{n}+1} =1$ always holds.

	\subsubsection{Comparison with the WCWL scheme in \cite{WCWL}} 
	Let $K=q^nL$ and $t=q^n-2^n\lfloor\frac{q-1}{2}\rfloor^{n}$ in \cite{WCWL}. 
	We compare the WCWL scheme with the proposed scheme in the following cases:
	\begin{align*}
		g_{\text{WCWL}}&=2\lfloor\frac{K}{K-tL+1}\rfloor
		=2\left\lfloor \frac{q^nL}{2^n\lfloor\frac{q-1}{2}\rfloor^{n}L+1} \right\rfloor.\\
	\end{align*}
	By Theorem~\ref{th:Lagrange}, we have the following result.
	\begin{align}\label{wcwlg}
		\frac{g_{\text{WCWL}}}{g}&=\frac{2\left\lfloor \frac{q^nL}{2^n\lfloor\frac{q-1}{2}\rfloor^{n}L+1} \right\rfloor}{2^n}=2^{1-n}\left\lfloor \frac{q^nL}{2^n\lfloor\frac{q-1}{2}\rfloor^{n}L+1} \right\rfloor
	\end{align}
	The ratio of our scheme’s coded caching gain to that of the WCWL scheme is given by $2^{1-n}\left\lfloor \frac{q^nL}{\lfloor q-1 \rfloor^n L + 1} \right\rfloor$. Since $q \geq 3$
	, this ratio will always be $\leq 1$, our scheme has been proven to superior to or equal to the WCWL scheme in terms of coded caching gain.
	\begin{itemize}
		\item If $(K-tL+1)|K$ or $K-tL=1$, we have 
		\begin{equation*}
			F_{\text{WCWL}}=K=q^nL, R_{\text{WCWL}}=\frac{(K-tL)(K-tL+1)}{2K}
			=\frac{2^n\lfloor\frac{q-1}{2}\rfloor^{n}\left(2^n\lfloor\frac{q-1}{2}\rfloor^{n}L+1\right)}{2q^n}.
		\end{equation*}
		By Theorem~\ref{th:Lagrange}, we have the following result.
		\begin{align*}
			\frac{F_{\text{WCWL}}}{F}=\frac{q^nL}{q^nL}=1, \quad
			\frac{R_{\text{WCWL}}}{R}=\frac{\frac{2^n\lfloor\frac{q-1}{2}\rfloor^{n}\left(2^n\lfloor\frac{q-1}{2}\rfloor^{n}L+1\right)}{2q^n}}{\lfloor\frac{q-1}{2}\rfloor^{n}L}=2^n
			\frac{2^n\lfloor\frac{q-1}{2}\rfloor^{n}L+1}{2q^nL} \textgreater \frac{2^{n}(q-1)^{n}}{2q^n}=\frac{1}{2}\cdot(2-\frac{2}{q})^n.
		\end{align*}

		\item If $\left \langle K \right \rangle_{K-tL+1}=K-tL$, we can get the WCWL scheme with
		\begin{equation*}
			F_{\text{WCWL}}=\left(2\lfloor\frac{K}{K-tL+1}\rfloor+1\right)K,\ \ \ R_{\text{WCWL}}=\frac{K-tL}{2\lfloor\frac{K}{K-t+1}\rfloor+1}
			=\frac{2^n\lfloor\frac{q-1}{2}\rfloor^{n}L}{2\left\lfloor \frac{q^nL}{2^n\lfloor\frac{q-1}{2}\rfloor^{n}L+1} \right\rfloor+1}.
		\end{equation*}
		By Theorem~\ref{th:Lagrange}, we can obtain the following.
		
		\begin{align*}
			\frac{F_{\text{WCWL}}}{F}&=\frac{(2\lfloor \frac{K}{K-t+1}\rfloor+1) K}{q^nL}=2\left\lfloor \frac{K}{K-t+1}\right\rfloor+1,\\
			\frac{R_{\text{WCWL}}}{R}&=\frac{\frac{2^n\lfloor\frac{q-1}{2}\rfloor^{n}L}{2\left\lfloor\frac{q^nL}{2^n\lfloor\frac{q-1}{2}\rfloor^{n}L+1}\right\rfloor+1}}{\lfloor\frac{q-1}{2}\rfloor^{n}L}
			=\frac{2^n}{2\left\lfloor \frac{q^nL}{2^n\lfloor\frac{q-1}{2}\rfloor^{n}L+1} \right\rfloor+1}> \frac{2^n(q-3)^nL}{2q^nL+(q-1)^nL+1}> \frac{2^n(q-3)^n}{3q^n}= \frac{1}{3}\left(2-\frac{1}{6}\right)^n.\\
		\end{align*}
		
		\item Otherwise, the WCWL scheme  can be obtain as follow.
		$$F_{\text{WCWL}}=2\left\lfloor\frac{K}{K-tL+1}\right\rfloor K,\ \ \ R_{\text{WCWL}}=\frac{K-tL}{2\lfloor\frac{K}{K-tL+1}\rfloor}
		={\frac{2^n\lfloor\frac{q-1}{2}\rfloor^{n}L}{2\left\lfloor \frac{q^nL}{2^n\lfloor\frac{q-1}{2}\rfloor^{n}L+1} \right\rfloor}}.$$ By Theorem~\ref{th:Lagrange}, we have
		\begin{align*}
			\frac{F_{\text{WCWL}}}{F}=\frac{2\left\lfloor\frac{K}{K-tL+1}\right\rfloor K}{q^nL}=2\left\lfloor \frac{K}{K-tL+1}\right\rfloor, 
			\frac{R_{\text{WCWL}}}{R}=\frac{\frac{2^n\lfloor\frac{q-1}{2}\rfloor^{n}L}{2\left\lfloor \frac{q^nL}{2^n\lfloor\frac{q-1}{2}\rfloor^{n}L+1} \right\rfloor}}{\lfloor\frac{q-1}{2}\rfloor^{n}L}=\frac{2^n}{2\left\lfloor \frac{q^nL}{2^n\lfloor\frac{q-1}{2}\rfloor^{n}L+1} \right\rfloor}>\frac{2^n}{2(\frac{q}{{q-1}})^{n}}=\frac{1}{2}\cdot(2-\frac{2}{q})^n.\\
		\end{align*}   
	\end{itemize}

	As a result, compared to the WCWL scheme in \cite{WCWL},  our scheme obtaining the  same subpacketization or the reduction amount of our subpacketization is larger than $2$ times. The multiplicative reduction amount by our transmission is at least $\frac{1}{3}\left(2-\frac{1}{6}\right)^n$. When $q>6$ and $n\geq\frac{\ln3}{\ln(2-\frac{6}{q})}$, the formula $\frac{1}{3}\left(2-\frac{1}{6}\right)^n \geq1$ always holds. The coded caching gain of our scheme is $2^{1-n}\left\lfloor \frac{q^nL}{2^n\lfloor\frac{q-1}{2}\rfloor^{n}L+1} \right\rfloor$ times that of the WCWL scheme. Since $K\geq 3^nL$, $2^{1-n}\left\lfloor \frac{q^nL}{2^n\lfloor\frac{q-1}{2}\rfloor^{n}L+1} \right\rfloor$ is always less than or equal to 1.
	\subsubsection{Comparison with the SR2 scheme in \cite{SR}} 
	Let $K=q^nL$ and $k=q^n-2^n\lfloor\frac{q-1}{2}\rfloor^{n}$ in \cite{SR}, i.e., the $10$th row of Table~\ref{tab-knowschemes}. 
	We compare the SR2 scheme with the proposed scheme in the following cases:
	
	\begin{align*}
		F_{\text{SR2}}=K=q^nL,\quad R_{\text{SR2}}=\frac{(K-kL)(K-kL+k)}{2K}
		=\frac{2^n\lfloor\frac{q-1}{2}\rfloor^{n} \left(2^n\lfloor\frac{q-1}{2}\rfloor^{n}L+q^n-2^n\lfloor\frac{q-1}{2}\rfloor^{n}\right)}{2q^n}.
	\end{align*}
	By Theorem~\ref{th:Lagrange}, we have the following result.
	\begin{align*}
		\frac{F_{\text{SR2}}}{F}=\frac{q^nL}{q^nL}=1, \quad
		\frac{R_{\text{SR2}}}{R}=\frac{\frac{2^n\lfloor\frac{q-1}{2}\rfloor^{n} \left(2^n\lfloor\frac{q-1}{2}\rfloor^{n}L+q^n-2^n\lfloor\frac{q-1}{2}\rfloor^{n}\right)}{2q^n}}{\lfloor\frac{q-1}{2}\rfloor^{n}L}=\frac{2^n2^n\lfloor\frac{q-1}{2}\rfloor^{n}(L-1)+q^n}{2q^nL}>\frac{1}{2L}\left(2-\frac{6}{q}\right).
	\end{align*}
	As a result, compared to the SR2 scheme in \cite{SR},  our scheme either obtaining the  same subpacketization. The multiplicative reduction amount by our transmission is at least $\frac{1}{2L}\left(2-\frac{6}{q}\right)$. When $q>6$ and $n\geq\frac{\ln2L}{\ln(2-\frac{6}{q})}$ ,the formula $\frac{1}{2L}\left(2-\frac{6}{q}\right)\geq1$ always holds.
	
	\subsection{Numerical comparisons} 
	\subsubsection{Comparison with the schemes in \cite{HKD,RK,WCWL,CWLZC,RK2021}}
\begin{figure}[htbp!]
	\centering
	\begin{minipage}[t]{0.45\textwidth}
		\centering
		\includegraphics[width=\linewidth]{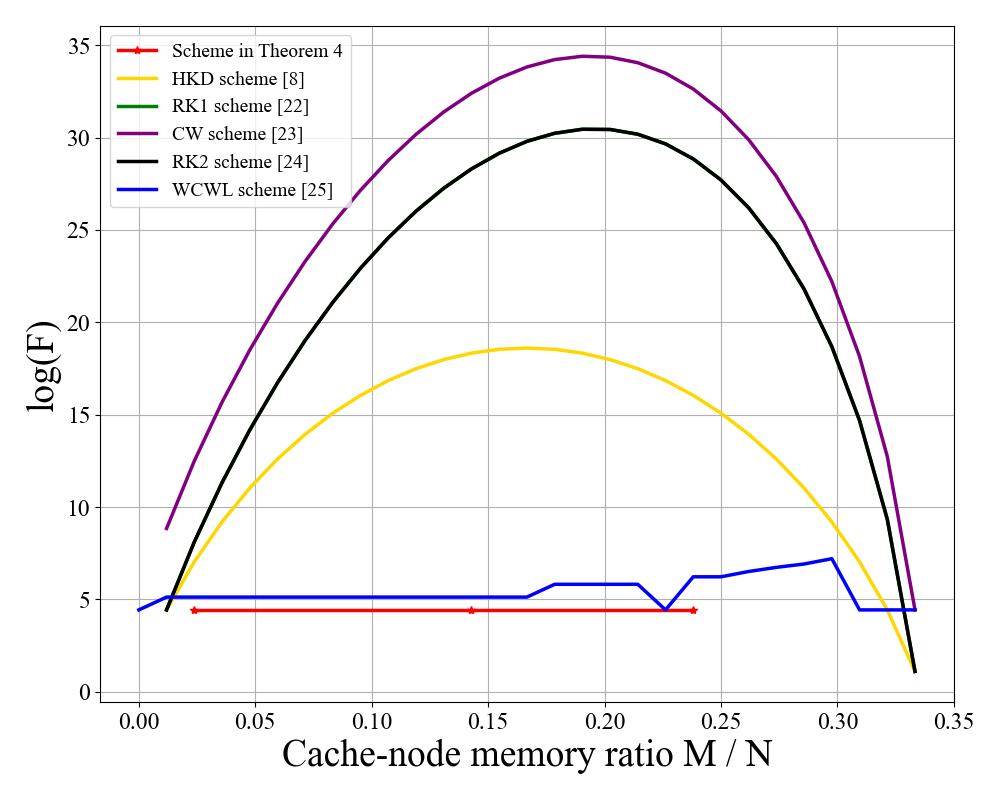}
		\caption{Memory ratio-subpacketization tradeoff for $K = 84$, $L = 3$}
		\label{Compare_t_F}
	\end{minipage}
	\hfill  
	\begin{minipage}[t]{0.45\textwidth}
		\centering
		\includegraphics[width=\linewidth]{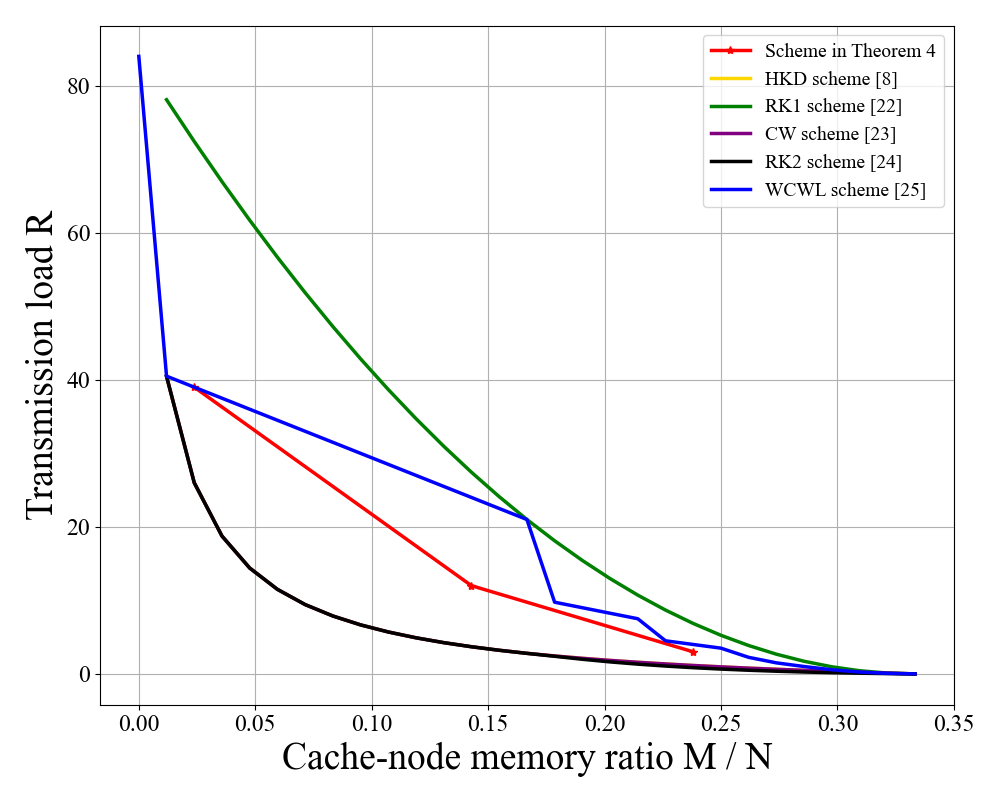}
		\caption{Memory ratio-load tradeoff for $K = 84$, $L = 3$}
		\label{Compare_t_R}
	\end{minipage}
\end{figure}

	When $K=84$ and $L=3$, we have our scheme and the schemes in \cite{HKD,RK,WCWL,CWLZC,RK2021} are shown in Fig.~\ref{Compare_t_F} and Fig.~\ref{Compare_t_R}. It can be seen that our scheme can achieve a significant lower subpacketization and a slightly higher load than the schemes in \cite{HKD,CWLZC,RK2021,RK} with exponential subpaketization, a lower subpacketization and a same or lower load than the scheme in \cite{WCWL}. For instance, when $M/N=0.14$, we have our scheme with $F=84$ and $R=11.40$; the schemes in \cite{HKD,CWLZC,RK,RK2021} have lowest subpacketization is $F \approx 8.687\times{10}^7>84$ and lowest load is $R = 4.10<11.40$. The WCWL scheme in \cite{WCWL} with $F\approx170>84$ and $R\approx27.98>11.40$.

	\subsubsection{Comparison with the schemes in ~\cite{SPE,SR_arX2021,SR,MR,NR}}
	We compared our proposed scheme in Theorem~\ref{th:Lagrange} with the schemes in \cite{SPE}, \cite{SR_arX2021}, \cite{SR}, \cite{MR}, and \cite{NR} by selecting specific user numbers and memory ratios, as listed in Table~\ref{tab-numerical-1}. Compared with the SPE scheme in \cite{SPE}, under similar numbers of users and memory ratios, our scheme achieves significantly lower subpacketization at the cost of a slightly higher transmission load. Compared with the SR1 scheme in \cite{SR_arX2021}, for the same number of users and memory ratio, the proposed scheme achieves a significantly lower transmission load while maintaining smaller subpacketization. For example, when $M/N = 0.0222$ and $K = 45$, the transmission load is reduced from $20$ to $5$, and the subpacketization is reduced from $90$ to $45$. Compared with the SR2 scheme in \cite{SR}, the proposed scheme achieves a lower transmission load and the same subpacketization level while having a close number of users and a close memory ratio. For example, when $K = 576$ and $M/N \approx 0.063$, the load decreases from $63$ to $9$ while keeping the same subpacketization. Compared with the MR scheme in \cite{MR} and the NR scheme in \cite{NR}, for the same number of users and memory ratio, the proposed scheme always achieves a lower or equal transmission load while maintaining the same subpacketization. For example, when $M/N = 0.1852$ and $K = 27$, the proposed scheme achieves a load of $R = 3$, which is lower than the $R = 5.3333$ achieved by the NR scheme.
	\begin{table}[htbp!]
		\centering
		\caption{The numerical comparison between  the scheme in Theorem~\ref{th:Lagrange} and the schemes in \cite{SPE,SR_arX2021,SR,MR,NR}}
		\label{tab-numerical-1}
		\begin{tabular}{|c|c|c|c|c|c|}
			\hline
			$K$   & $M/N$   & Scheme  & Parameters & Load   & Subpacketization \\ \hline 
			$28$ & $0.0714$ & SPE scheme in  \cite{SPE} & $z=3$ & $6.7500$ & $168$   \\ 
			$27$ & $0.0370$ & Scheme in Theorem~\ref{th:Lagrange}  & $(v,n,L)=(27,1,3)$ & $12$ & $27$  \\ \hline
			$64$ & $0.0313$ & SPE scheme in  \cite{SPE}  & $z=7$ & $14.8846$ & $832$  \\
			$63$ & $0.0159$ & Scheme in Theorem~\ref{th:Lagrange} & $(v,n,L)=(63,1,7)$ & $28$ & $63$  \\  \hline
			$288$ & $0.0069$ & SPE scheme in  \cite{SPE}  & $z=7$ & $89.3841$ & $19872$  \\
			$287$ & $0.0035$ & Scheme in Theorem~\ref{th:Lagrange}  & $(v,n,L)=(287,1,7)$ & $140$ & $287$  \\  \hline
			$1112$ & $0.0018$ & SPE scheme in  \cite{SPE}  & $z=11$ & $360.0311$ & $303576$  \\
			$1111$ & $0.0009$ & Scheme in Theorem~\ref{th:Lagrange}  & $(v,n,L)=(1111,1,11)$ & $550$ & $1111$  \\  \hline
			
			$45$ & $0.0222$ & SR1 scheme in  \cite{SR_arX2021} & $(k,z)=(1,5)$ & $20$ & $90$  \\
			$45$ & $0.0222$ & Scheme in Theorem~\ref{th:Lagrange}& $(v,n,L)=(45,2,5)$ & $5$ & $45$  \\ \hline
			$147$ & $0.0884$ & SR1 scheme in  \cite{SR_arX2021} & $(k,z)=(13,3)$ & $54$ & $294$  \\
			$147$ & $0.0884$ & Scheme in Theorem~\ref{th:Lagrange}& $(v,n,L)=(147,2,3)$ & $27$ & $147$  \\ \hline
			$245$ & $0.0531$ & SR1 scheme in  \cite{SR_arX2021} & $(k,z)=(13,5)$ & $90$ & $490$  \\
			$245$ & $0.0531$ & Scheme in Theorem~\ref{th:Lagrange}& $(v,n,L)=(245,2,5)$ & $45$ & $245$  \\ \hline
			$1331$ & $0.0158$ & SR1 scheme in  \cite{SR_arX2021} & $(k,z)=(21,11)$ & $550$ & $2662$  \\
			$1331$ & $0.0158$ & Scheme in Theorem~\ref{th:Lagrange}& $(v,n,L)=(1331,2,11)$ & $275$ & $1331$  \\ \hline
			
			$80$ & $0.1250$ & SR2 scheme in  \cite{SR} & $(k,L)=(10,5)$ & $7.5$ & $80$  \\
			$80$ & $0.0875$ & Scheme in Theorem~\ref{th:Lagrange}& $(v,n,L)=(80,2,5)$ & $5$ & $80$  \\ \hline
			$144$ & $0.0625$ & SR2 scheme in  \cite{SR} & $(k,L)=(9,9)$ & $15.75$ & $144$  \\
			$144$ & $0.0486$ & Scheme in Theorem~\ref{th:Lagrange}& $(v,n,L)=(144,2,9)$ & $9$ & $144$  \\ \hline
			$576$ & $0.0625$ & SR2 scheme in  \cite{SR} & $(k,L)=(36,9)$ & $63$ & $576$  \\
			$576$ & $0.0642$ & Scheme in Theorem~\ref{th:Lagrange}& $(v,n,L)=(576,3,9)$ & $9$ & $576$  \\ \hline
			$996$ & $0.1667$ & SR2 scheme in  \cite{SR} & $(k,L)=(166,5)$ & $27.6667$ & $996$  \\
			$1215$ & $0.1737$ & Scheme in Theorem~\ref{th:Lagrange}& $(v,n,L)=(1215,5,5)$ & $5$ & $1215$  \\ \hline
			
			$27$ & $0.1852$ & MR scheme in  \cite{MR} & $(i,L)=(5,3)$ & $3$ & $27$  \\
			$27$ & $0.1852$ & Scheme in Theorem~\ref{th:Lagrange}  & $(v,n,L)=(27,2,3)$ & $3$ & $27$  \\  \hline
			$45$ & $0.0222$ & MR scheme in  \cite{MR} & $(i,L)=(1,5)$ & $20$ & $45$  \\
			$45$ & $0.0222$ & Scheme in Theorem~\ref{th:Lagrange}& $(v,n,L)=(45,2,5)$ & $5$ & $45$  \\ \hline
			$245$ & $0.0531$ & MR scheme in  \cite{MR} & $(i,L)=(13,5)$ & $90$ & $245$  \\
			$245$ & $0.0531$ & Scheme in Theorem~\ref{th:Lagrange}& $(v,n,L)=(245,2,5)$ & $45$ & $245$  \\ \hline
			$3125$ & $0.1181$ & MR scheme in  \cite{MR} & $(i,L)=(369,5)$ & $320$ & $3125$  \\
			$3125$ & $0.1181$ & Scheme in Theorem~\ref{th:Lagrange}  & $(v,n,L)=(3125,4,5)$ & $80$ & $3125$  \\  \hline
			
			$27$ & $0.1852$ & NR scheme in  \cite{NR} & $(i,L)=(5,3)$ & $5.3333$ & $27$  \\
			$27$ & $0.1852$ & Scheme in Theorem~\ref{th:Lagrange}  & $(v,n,L)=(27,2,3)$ & $3$ & $27$  \\  \hline
			$45$ & $0.0222$ & NR scheme in  \cite{NR} & $(i,L)=(1,5)$ & $35.5556$ & $45$  \\
			$45$ & $0.0222$ & Scheme in Theorem~\ref{th:Lagrange}& $(v,n,L)=(45,2,5)$ & $5$ & $45$  \\ \hline
			$245$ & $0.0531$ & NR scheme in  \cite{NR} & $(i,L)=(13,5)$ & $132.2449$ & $245$  \\
			$245$ & $0.0531$ & Scheme in Theorem~\ref{th:Lagrange}& $(v,n,L)=(245,2,5)$ & $45$ & $245$  \\ \hline
			$1331$ & $0.0158$ & NR scheme in  \cite{NR} & $(i,L)=(21,11)$ & $909.0909$ & $1331$  \\
			$1331$ & $0.0158$ & Scheme in Theorem~\ref{th:Lagrange}& $(v,n,L)=(1331,2,11)$ & $275$ & $1331$  \\ \hline
			
		\end{tabular}
	\end{table}
	
	\section{Conclusion}\label{sec-conclu}
	This paper proposes a novel combinational structure called as cyclic multi-access non-half-sum disjoint packing (CMA-NHSDP), which is an extension of the traditional NHSDP structure to MACC system. The CMA-NHSDP can realize a multi-access coded caching scheme with linear subpacketization while minimizing transmission load. In terms of performance comparison, compared to some existing schemes with linear subpacketization, our scheme has larger coded caching  gain, thus achieving a lower transmission load. And our scheme achieves far lower subpacketization and slightly higher load compare to some existing schemes with exponential subpacketization.

	\appendices
	
	\section{Proof of Theorem~\ref{th-main-PDA}
	}
	\label{sec:proof of thm2}
	Consider the first condition in Definition \ref{def-NHSDP}. Let ${\bf a}=(a_1,a_2,\ldots,a_n)$, ${\bf a}'=(a'_1,a'_2,\ldots,a'_n)$, ${\bf \alpha}=(\alpha_1,\alpha_2,\ldots,\alpha_n)$, and  ${\bf\alpha}'=(\alpha'_1,\alpha'_2, \ldots,\alpha'_n)$ are four vectors, where ${\bf a}, {\bf a} '\in \mathcal{A}$ and ${\bf\alpha},{\bf\alpha}' \in \{1,-1\}^n$.  From \eqref{eq-family} we have two integers of $\mathfrak{D}$, i.e., 
	\begin{equation}
		\begin{aligned}
			x= \alpha_1 (a_1 x_1 + g(1)) + \cdots + \alpha_n (a_n x_n + g(n)) \in \mathcal{D}_{\mathbf{a}} \quad
			\text{and} \quad
			y= \alpha_1' (a_1' x_1 + g(1)) + \cdots + \alpha_n' (a_n' x_n + g(n)) \in \mathcal{D}_{\mathbf{a}}
		\end{aligned}
	\end{equation} Since
	$v\geq 2\phi(m_1,m_2,\ldots,m_n)+1=2\sum_{i=1}^{n}\left( f(i) + g(i) \right)+1$, both $\sum_{i=1}^{n}\left(a_i x_i + g(i)\right)$ and  $\sum_{i=1}^{n}\left(a_i' x_i + g(i)\right)$ are less than $\frac{v-1}{2}$. In addition, from \eqref{eq-RF} for each $i\in[n-1]$ we have $x_{i+1}>2\sum_{j=1}^{i}\left( f(j) + g(j) \right)$ which implies that
	\begin{align}
		\label{eq-relation}
		a_{i+1}x_{i+1}+g(i+1)>2\sum_{j=1}^{i}\left(a_j x_j + g(j)\right).
	\end{align} If $x=y$, we have $\alpha_n=\alpha'_n$ and $a_n=a'_n$. Otherwise, from \eqref{eq-relation} with $i=n-1$, if $a_n\neq a'_n$ we have $x\neq y$. Furthermore, if $\alpha_n\neq \alpha'_n$, without loss of generality, we assume that $\alpha_n<0$ and $\alpha'_n>0$. Then we have $\frac{v-1}{2}< x<v$ and $y\leq \frac{v-1}{2}$, which implies $x\neq y$. This contradicts our hypothesis that $x=y$. So we only need to consider the case $x-\alpha_n(a_n x_n + g(n))=y-\alpha'_n(a'_n x_n + g(n))$. Similarly we can obtain $a_{n-1}=a'_{n-1}$ and $\alpha_{n-1}=\alpha'_{n-1}$ from \eqref{eq-relation} with $i=n-2$.
	Using the aforementioned proof method, we can analogously obtain $a_i=a'_i$ and $\alpha_i=\alpha'_i$, which implies  $\alpha_i\left(a_ix_i+g(i)\right)=\alpha'_i\left(a'_ix_i+g(i)\right)$ for each integer $i\in[n]$. Then ${\bf a}={\bf a}'$ and ${\bf \alpha}={\bf \alpha}'$. So each integer in $\mathfrak{D}$ appears exactly once.

	Now let us check the property of non-half-sum. For any vector ${\bf a}=(a_1,a_2,\ldots,a_n)\in \mathcal{A}$ and any two different vectors ${\bf \alpha}=(\alpha_1,\alpha_2,
	\ldots,\alpha_n)$, ${\bf \alpha}'=(\alpha'_1,\alpha'_2, \ldots,\alpha'_n)\in\{-1,1\}^n$, let us consider the half-sum of integers  
	\begin{equation}
		\begin{aligned}
			x = \alpha_1 (a_1 x_1 + g(1)) + \cdots + \alpha_n (a_n x_n + g(n)) \in \mathcal{D}_{\mathbf{a}} \quad \text{and} \quad
			y = \alpha_1' (a_1 x_1 + g(1)) + \cdots + \alpha_n' (a_n x_n + g(n)) \in \mathcal{D}_{\mathbf{a}}
		\end{aligned}
	\end{equation}i.e., $$\frac{x+y}{2}=\frac{\alpha_1+\alpha'_1}{2}\cdot (a_1 x_1 + g(1))+\cdots+\frac{\alpha_n+\alpha'_n}{2}\cdot (a_n x_n + g(n)).$$ 
	By our hypothesis ${\bf \alpha}\neq {\bf \alpha}'$ we have $x\neq y$. In addition, since $\alpha_1,\alpha_2\in \{-1,1\}$ it follows that $\frac{\alpha_i+\alpha'_i}{2}\in\{-1,0,1\}$ for each $i\in[n]$. Furthermore there exists at least one integer $i'\in[n]$ such that $\frac{\alpha_i+\alpha'_i}{2}=0$. Otherwise we have ${\bf \alpha}={\bf \alpha}'$ which implies $x=y$. This contradicts our hypothesis $x\neq y$. In the following we will show that $\frac{x+y}{2}$ does not occur in $\mathfrak{D}$. Assume that there exists two vectors ${\bf a}'=(a'_1,a'_2,\ldots,a'_n)\in \mathcal{A}$ and ${\bf \beta}=(\beta_1,\beta_2,\ldots,\beta_n)\in\{-1,1\}^n$ such that  
	\begin{equation}
		\begin{aligned}
			z &= a'_1\left(\beta_1x_1 + g(1)\right) + \cdots + a'_n\left(\beta_nx_n + g(n)\right) \\
			&= \frac{\alpha_1 + \alpha'_1}{2} \cdot a_1x_1 + \cdots + \frac{\alpha_n + \alpha'_n}{2} \cdot a_nx_n.
		\end{aligned}
	\end{equation}Similar to the proof of the uniqueness of each integer in $\mathfrak{D}$ introduced above, we can get $\beta_i=\frac{\alpha_i+\alpha'_i}{2}$ and $a'_i=a_i$ for every $i\in [n]$. Then there exists at  least one $\beta_{i'}=0$ for some $i'\in[n]$. This contradicts our definition rule given in \eqref{eq-family}, that is, $\beta_{i'} \in \{-1,1\}^n$. So the half-sum of any two different integers in  $\mathcal{D} \in \mathfrak{D}$ does not occur in  $\mathscr{D}$.
	
	Finally, we verify the $L$-consecutive condition. 
	Recall that every element in $\mathscr{D}$ can be represented as $\sum_{j=i+1}^{n} \alpha_j\left( a_jx_j + g(j) \right)+\alpha_i \left( a_ix_i + g(i) \right)+\sum_{h=1}^{i-1} \alpha_h\left( a_hx_h + g(h) \right)$, where $i \in [n]$, $\alpha_i \in \{-1,1\}$, $a_i \in [m_i]$.
	To show that $\mathbb{Z}_v \setminus \mathcal{D}$ can be partitioned into disjoint subsets of $L$ consecutive integers, it suffices to prove that for any two neighboring elements in the sorted order of $\mathcal{D}$, their difference is equal to $1$ modulo $L$. We analyze the possible patterns of the coefficients. Due to the recursive construction of $\mathcal{D}$, the neighboring elements can be categorized into three cases.
	\begin{itemize}
		\item \textbf{Case 1:} When $ \alpha_i =-1$, $a_i>1$, $\alpha_h=1$, and $a_h=m_h$, i.e.,$\sum_{j=i+1}^{n} \alpha_j\left( a_jx_j + g(j) \right)- \left( a_ix_i + g(i) \right)+\sum_{h=1}^{i-1} \left( m_hx_h + g(h) \right)$, the neighboring larger element is $\sum_{j=i+1}^{n} \alpha_j\left( a_jx_j + g(j) \right)- \left( (a_i-1)x_i + g(i) \right)-\sum_{h=1}^{i-1} \left(m_hx_h + g(h) \right)$, and the difference is $x_i- 2\sum_{h=1}^{i-1} \left( m_hx_h + g(h) \right)=1$;
		\item \textbf{Case 2:} When $ \alpha_i =-1$, $a_i=1$, $\alpha_h=1$, and $a_h=m_h$, i.e., $\sum_{j=i+1}^{n} \alpha_j\left( a_jx_j + g(j) \right)- \left(x_i + g(i) \right)+\sum_{h=1}^{i-1} \left(m_hx_h + g(h) \right)$, the neighboring larger element is $\sum_{j=i+1}^{n} \alpha_j\left( a_jx_j + g(j) \right)+ \left( x_i + g(i) \right)-\sum_{h=1}^{i-1} \left(m_hx_h + g(h) \right)$, and the difference is $2\left(x_i + g(i)-\sum_{h=1}^{i-1} \left(m_hx_h  + g(h) \right)\right)$;
		Let us consider the difference under modulo $L$\\
		When $i=1$,\\
		\begin{equation}
			\begin{aligned}
				\left\langle 2\left( x_i + g(i) - \sum_{h=1}^{i-1} \left( m_hx_h + g(h) \right) \right) \right\rangle_L
				= \left\langle 2\left( x_1 + g(1) \right) \right\rangle_L
				= \left\langle 2\left( 1 + \frac{L-1}{2} \right) \right\rangle_L = 1.
			\end{aligned}
		\end{equation}
		When $i\geq2$,\\
		\begin{equation}
			\begin{aligned}
				\left\langle 2\left( x_i + g(i) - \sum_{h=1}^{i-1} \left( m_hx_h + g(h) \right) \right) \right\rangle_L
				&= \left\langle 2\left( 2\sum_{h=1}^{i-1} \left( m_hx_h + g(h) \right) + 1 + g(i) - \sum_{h=1}^{i-1} \left( m_hx_h + g(h) \right) \right) \right\rangle_L \\
				&= \left\langle 2\left( \sum_{h=1}^{i-1} \left( m_hx_h + g(h) \right) + 1 + g(i) \right) \right\rangle_L \\
				&= \left\langle 2\left( \sum_{h=2}^{i} \left( m_{h-1}x_{h-1} + g(h) \right) + 1 + g(1) \right) \right\rangle_L \\
				&= \left\langle 2\left( g(1) + 1 \right) \right\rangle_L = 1.
			\end{aligned}
		\end{equation}
	\item \textbf{Case 3:} When $ \alpha_i =1$, $\alpha_h=1$, and $a_h=m_h$, i.e.,$\sum_{j=i+1}^{n} \alpha_j\left( a_jx_j + g(j) \right)+ a_ix_i + g(i)+\sum_{h=1}^{i-1} \left( m_hx_h + g(h) \right)$, the neighboring larger element is $\sum_{j=i+1}^{n} \alpha_j\left( a_jx_j + g(j) \right)+ \left( (a_i+1)x_i + g(i) \right)-\sum_{h=1}^{i-1} \left(m_hx_h + g(h) \right)$, and the difference is $x_i- 2\sum_{h=1}^{i-1} \left( m_hx_h + g(h) \right)=1$;
	\end{itemize}

	Above all, in the sorted $\mathscr{D}$,  the difference between any two neighboring elements is calculated. Taking the modulo $L$ operation on this difference yields a result of 1. So partition the complement of $\mathscr{D}$ in $\mathbb{Z}_v$ into disjoint groups, each of which contains exactly $L$ consecutive elements from $\mathbb{Z}_v$. Then the proof is completed.
	
	\bibliographystyle{IEEEtran}
	\bibliography{reference}
\end{document}